\documentclass[aps,twocolumn,showpacs,preprintnumbers,superscriptaddress,longbibliography]{revtex4-1}

\usepackage{graphicx}  
\usepackage{subfigure}
\usepackage{multirow}

\usepackage{fancyhdr}
\usepackage{longtable}
\usepackage{parskip}
\usepackage[T1]{fontenc}
\usepackage{dcolumn}   
\usepackage{afterpage}
\usepackage{bm}        
\usepackage{amsfonts}  
\usepackage{amsmath}   
\usepackage{amssymb}   
\usepackage{physics}
\usepackage{svg}
\usepackage{hyperref}
\usepackage{mathtools}
\usepackage{float}
\usepackage{comment}
\usepackage{xcolor}

\AtBeginDocument{%
    \newwrite\bibnotes
    \def\bibnotesext{Notes.bib}
    \immediate\openout\bibnotes=\jobname\bibnotesext
    \immediate\write\bibnotes{@CONTROL{REVTEX41Control}}
    \immediate\write\bibnotes{@CONTROL{%
    apsrev41Control,author="08",editor="1",pages="1",title="0",year="1"}}
     \if@filesw
     \immediate\write\@auxout{\string\citation{apsrev41Control}}%
    \fi
}%
\DeclareUnicodeCharacter{0308}{HERE!HERE!}

\newcommand{\pwisein}{\left\{ \begin{array}{ll}}
\newcommand{\pwiseout}{\end{array}\right.}

\usepackage{indentfirst}
\begin{document}

\title{Wigner-function formalism for the detection of single microwave pulses in a resonator-coupled double quantum dot}

\author{Drilon Zenelaj}

\affiliation{\it Physics Department and NanoLund, Lund University, Box 118, 22100 Lund, Sweden}

\author{Peter Samuelsson}

\affiliation{\it Physics Department and NanoLund, Lund University, Box 118, 22100 Lund, Sweden}

\author{Patrick P. Potts}

\affiliation{\it Department of Physics and Swiss Nanoscience Institute, University of Basel, Klingelbergstrasse 82, 4056 Basel, Switzerland}

\date{\today}

\begin{abstract}  
Semiconductor double quantum dots (DQD) coupled to superconducting microwave resonators offer a promising platform for the detection of single microwave photons. In previous works, the photodetection was studied for a monochromatic source of microwave photons. Here, we theoretically analyze the photodetection of single microwave pulses. The photodetection in this case can be seen as a non-linear filtering process of an incoming signal, the pulse, to an outgoing one, the photocurrent. This analogy to signal processing motivated the derivation of a Wigner-function formalism which provides a compelling visualization of the time and frequency properties of the photodetector for low intensities. We find a trade-off between detecting the time and the frequency of the incoming photons in agreement with the time-energy uncertainty relation. As the intensity of the source increases, the photodetection is influenced by coherent Rabi oscillations of the DQD. Our findings give insight into the time-dependent properties of microwave photons interacting with electrons in a DQD-resonator hybrid system and provide guidance for experiments on single microwave pulse detection.

\end{abstract}

\maketitle 

\section{Introduction}
The ability to efficiently and continuously detect single photons \cite{eisaman} is a key prerequisite in a variety of emergent quantum technologies, such as quantum cryptography and key distribution \cite{pirandola}, quantum random number generation \cite{herrero}, and linear optics quantum computation  \cite{knill}. Most developments in the field of single-photon detectors, like near-unit photodetection efficiency \cite{pernice} and photon-number resolved detection \cite{kardynal}, have so far been achieved in the optical domain. In said frequency regime, standard semiconductor diodes are employed, where the bandgap energy of the semiconductor is on the order of the energy of the optical photons \cite{pearsall}.

This approach does, however, not work in the microwave domain, since microwave photons have an energy which is around five orders of magnitude smaller compared to optical photons. A different architecture is thus needed for the detection of single microwave photons. In recent years, different architectures have been investigated both theoretically and experimentally \cite{romero,chen,fan,sankar,kyriienko,gu,inomata,narla,opremcak,besse,kono,huard,essig,basset}. Detectors based on superconducting qubits \cite{narla,opremcak,huard,besse,kono,balembois,petrovnin} or SNS junctions \cite{basset} have demonstrated functionalities such as near-unit photodetection efficiency \cite{narla,opremcak,huard,basset} and quantum non-demolition measurements \cite{besse,kono} of itinerant photons. 

One promising candidate, the one relevant for this work, comprises a semiconductor double quantum dot (DQD) coupled to a driven superconducting microwave resonator. Recently, efficient and continuous microwave photodetection in said system has been achieved, with a photodetection reaching up to 25\% \cite{haldar3}, moving close to the theoretically predicted unit efficiency \cite{Wong}. The detection mechanism is an analogue to the semiconductor photodiode, as photons in the resonator generate an electrical photocurrent by exciting an electron from the ground to the excited state of the DQD.

So far, photodetection in the DQD-resonator system has been studied in the case where the drive is a monochromatic source of microwave radiation \cite{Khan,Zenelaj}. Extending on previous works on photodetection for a monochromatic drive, the focus of the present work is to investigate the time-dependent photodetector properties of the DQD-resonator system when subjected to an input of microwave pulses. Quite generally, the photodetection process constitutes a nonlinear filtering \cite{Cohen}, converting an incoming signal, the coherent drive pulse with complex amplitude $f(t)$, into a an outgoing signal, the (real) photocurrent $I(t)$. A compelling mathematical tool for the study of such a filtering process, in the weak drive limit, is the Wigner function \cite{mecklen}. This formalism gives access to study the time and frequency dependence of the detector and highlights the well-known time-frequency uncertainty relation, preventing the simultaneous measurement of a photon's arrival time and its frequency. We focus on two well-known and analytically tractable pulse shapes, Gaussian and Lorentzian. We characterize the performance of the photodetector by comparing the amplitude, shape and timing of the outgoing photocurrent pulse to the same properties of the incoming drive pulse. We find that the photodetector works best in the low-drive regime and when the bandwidth of the detector is large compared to the linewidth of the drive pulse. Moving away from the low-drive regime, we find that the photocurrent can exhibit oscillations in time, which we attribute to Rabi oscillations of the DQD. The findings of this work may encourage additional experiments on hybrid semiconductor-resonator photodetectors. Furthermore, they will establish a foundation for future theoretical studies on photodetection using various microwave sources, including non-classical ones, and for exploring functionalities such as single-photon and photon-number-resolved detection.

This paper is organized as follows. in Sec.~\ref{sandm}, we present our theoretical model for describing the DQD-resonator system and demonstrate its working principle as an efficient photodetector. Additionally, we review the necessary conditions to achieve near unit photon-to-electron conversion efficiency. In Sec.~\ref{wfdd}, we present the Wigner-function formalism for the drive and detector, respectively, and show how it can be used to compute the photocurrent. In Sec.~\ref{prdo}, we analyze the performance of the photodetector by employing several performance quantifiers. In Sec.~\ref{bld}, we analyze the photodetection away from the low-drive limit. We conclude and give a brief outlook in Sec.~\ref{cao}.

\section{System and model}\label{sandm}
\begin{figure}[ht!]
\centering
    \includegraphics[width=0.99\linewidth]{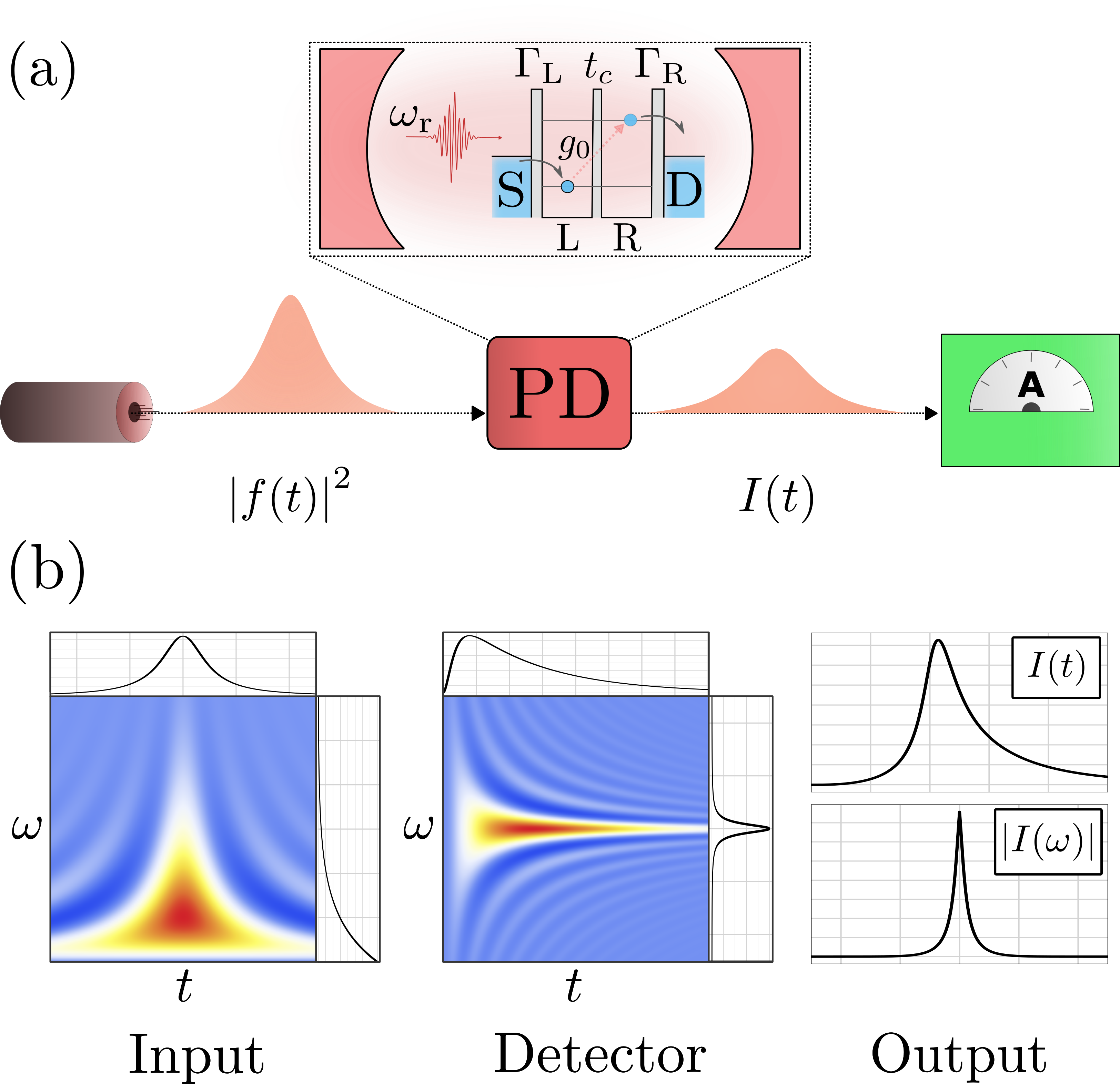}
    \caption{(a) A photodetector (PD) transforms an incoming microwave pulse to an outgoing electrical pulse. The PD consists of a double quantum dot (DQD) embedded in a microwave resonator with resonance frequency $\omega_\text{r}$. The DQD is tunnel coupled to a source (S) and drain (D) lead. An electron from S tunnels into the left dot L with the rate $\Gamma_\text{L}$. The electron can then absorb a photon through the coupling to the resonator with the bare coupling strength $g_0$ and move to the right dot R, where it can tunnel out to D with the rate $\Gamma_\text{R}$, producing an electrical photocurrent through the DQD. The incoming microwave pulse is described by a complex amplitude $f(t)$, the outgoing photocurrent by $I(t)$ (see main text). (b) Left and middle: Wigner function representation of the incoming drive for a Lorentzian pulse and the PD, respectively, and their marginals. Right: Outgoing photocurrent as a function of time (top panel) and frequency (bottom panel).}
    \label{fig:1.0}
\end{figure}

The photodetector system, shown in Fig.~\ref{fig:1.0} and studied in Refs.~\cite{Childress,Frey,Bergenfeldt,Xu,Lambert,Bergenfeldt2,Liu,Wong,Stockklauser,Khan,Zenelaj,Haldar2,Haldar}, consists of a double quantum dot (DQD) coupled to a microwave resonator. The DQD is further coupled to fermionic source and drain leads. In contrast to earlier works, we here consider the system driven by applying pulses of microwave photons - a coherent tone with a time-modulated amplitude.
The DQD acts as a detector for the microwave pulses by transforming them into electrical photocurrent pulses. 

Following closely the notation in Ref.~\cite{Zenelaj}, the Hamiltonian describing the driven DQD-resonator system in the frame rotating at the center frequency of the pulse $\omega_\text{l}$ is given by \cite{Faraon,Majumdar,Rasero}
\begin{equation}\label{H6}
\begin{split}
    H&=\Delta_\text{d}\frac{\sigma_z}{2}+\Delta_\text{r}a^\dagger a +g(a^\dagger \sigma_-+a\sigma_+)\\& + \sqrt{\kappa_\text{in}}\left[f^*(t)a  +f(t)a^\dagger\right],
\end{split}
\end{equation}
where $\Delta_\text{d}=\Omega-\omega_\text{l}$ ($\Delta_\text{r}=\omega_\text{r}-\omega_\text{l}$) gives the detuning between the DQD (resonator) and the drive, $\kappa_\text{in}$ denotes the coupling strength between the impinging photons and the resonator and $f(t)$ is the time-dependent amplitude of the pulse. Here, $\Omega=\sqrt{\epsilon^2+4t_c^2}$ is the splitting between the DQD ground and excited states, denoted $\ket{g}$ and $\ket{e}$, with $\epsilon$ being the energy detuning between quantum dot L and R and $t_c$ is the inter-dot tunnel coupling. The DQD operators include the Pauli $z$-matrix $\sigma_z=\ketbra{e}{e}-\ketbra{g}{g}$ as well as the raising and lowering operators $\sigma_+ =\ketbra{e}{g}$ and $\sigma_-=\sigma_+^\dagger=\ketbra{g}{e}$. We note that the state space of the DQD also includes $\ket{0}$, denoting the state with no electrons on the DQD, taken to have zero energy without loss of generality. The state $\ket{0}$ together with the ground state $\ket{g}$ and the excited state $\ket{e}$ span the Hilbert space of the DQD. The operators $a^\dagger$ ($a$) create (annihilate) photons in a resonator mode with frequency $\omega_\text{r}$. The coupling constant $g=g_0 \sin(\theta)$ between electrons and photons comes from the well-known Jaynes-Cummings model \cite{Cummings}, describing light-matter interaction on the quantum level, with a single photon mode coupled to a two-level quantum system. Here, $g_0$ is the bare coupling strength and the mixing angle $\theta$ is defined by $\cos(\theta)=-\epsilon/\Omega$. While originally applied to the interaction between atoms and photons, it has over the last two decades increasingly been used to describe the interaction between microwaves and solid state two-level systems, as described for DQDs in e.g. Ref.~\cite{Childress}, where also a derivation of $g$ is presented.

The rates for electron tunneling between the leads and the L and R dot are $\Gamma_\text L$ and $\Gamma_\text R$, respectively. Further, the source and drain leads are kept at the same potential and energies are counted with respect to the lead chemical potential. As illustrated in Fig.~\ref{fig:1.0}, the DQD ground and excited states energies are symmetric around zero. Assuming a small temperature $T$ of the leads, such that $k_\text B T \ll \Omega$, electrons from the leads can only enter the DQD into the ground state and leave from the excited state, with respective rates 
\begin{equation}\label{etunrat}
    \Gamma_{g0}=\Gamma_\text{L,in}+\Gamma_\text{R,in},\quad
    \Gamma_{0e}=\Gamma_\text{L,out}+\Gamma_\text{R,out},  
\end{equation}
where
\begin{equation}
\begin{split}
    &\Gamma_\text{L,in} = \Gamma_\text{L}\cos^2(\theta/2), \quad \Gamma_\text{R,in}=\Gamma_\text{R}\sin^2(\theta/2), \\
    &\Gamma_\text{L,out} = \Gamma_\text{L}\sin^2(\theta/2), \quad \Gamma_\text{R,out}=\Gamma_\text{R}\cos^2(\theta/2).
\end{split}    
\end{equation} 
The tunneling rates of Eq.~\eqref{etunrat} are derived within a standard framework for open quantum systems, starting with a microscopic Hamiltonian including both the DQD and the electronic reservoir. In addition, rotating from the local QD level, or charge state, basis to the ground-excited state basis, the mixing angle $\theta$ enters the tunnel rates.
In addition to electron tunneling, we account for pure dephasing of the DQD state with rate $\gamma_\phi$, relaxation of an electron from the excited to the ground state of the DQD with rate $\gamma_-$ and photonic losses with rate $\kappa$. The dynamics of the driven DQD-resonator system can then be described by a Lindblad master equation of the form \cite{Lindblad,Breuer}
\begin{equation}\label{LME}
\begin{split}
    &\partial_t \rho(t)=-i[H,\rho(t)]+\Gamma_{g0}\mathcal{D}[s_g^\dagger]\rho(t)+\Gamma_{0e}\mathcal{D}[s_e]\rho(t)\\ 
    &+\frac{\gamma_\phi}{2}\mathcal{D}[\sigma_z]\rho(t)+\gamma_-\mathcal{D}[\sigma_-]\rho(t)+\kappa\mathcal{D}[a]\rho(t),
\end{split}
\end{equation}
with the superoperator $\mathcal{D}[x]\rho(t)=x\rho(t)x^\dagger -\frac{1}{2}\{x^\dagger x,\rho(t)\}$, where $s_g^\dagger = \ketbra{g}{0}$ and $s_e=\ketbra{0}{e}$. Here, $\rho(t)$ is the density matrix of the joint DQD-resonator system, i.e., $\rho(t)$ lives in a Hilbert space spanned by the three DQD states $\ket{0},\ket{e},\ket{g}$ and the infinite states of the harmonic oscillator describing the resonator.
The electrical photocurrent $I(t)$ flowing through the DQD, the photocurrent, is given by 
\begin{equation}\label{pc}
    I(t)/e=\Gamma_\text{R,out} p_e-\Gamma_\text{R,in} p_0,
\end{equation}
evaluated at the lead coupled to the R dot. Here, $p_e=\langle e|\tilde \rho |e\rangle$ and $p_0=\langle 0|\tilde \rho |0\rangle$, with $\tilde \rho$ being the reduced density matrix of the DQD.

The photon-to-electron conversion efficiency, the probability that an incident photon is converted into an electron tunneling through the DQD, is a key figure of merit for the photodetector, as discussed further below. Connecting to previous works \cite{Wong,Khan,Zenelaj}, for a monochromatic drive at frequency $\omega_\text l$ with constant amplitude $\sqrt{\dot{N}}$, the efficiency is defined as 
\begin{equation}
    \eta_\text{mc}(\omega_\text l)=\frac{I_\text{mc}(\omega_\text l)/e}{\dot{N}},
\end{equation}
where $I_\text{mc}$ is the dc electrical photocurrent and $\dot{N}$ denotes the rate of impinging photons. In Refs.~\cite{Wong,Zenelaj} it was shown that ideal, unit-efficiency photodetection could be achieved for a monochromatic drive with the photodetector setup in Fig.~\ref{fig:1.0}. For completeness and further use below, we summarize the parameter conditions required for ideal photodetection in Tab.~\ref{tab:ideal}. In the table, $\tilde{\Gamma}=\Gamma_{0e}+\gamma_-+2\gamma_\phi$ denotes the total electron decoherence rate and $C = 4g^2/(\tilde{\Gamma}\kappa)$ the cooperativity. As can been seen from the table, a cooperativity equal to one is ideal for photodetection. We want to emphasize that this implies that an ideal photodetector does not require strong coupling between the DQD and the cavity.
\renewcommand{\arraystretch}{1.3}
\begin{table}[h]
    \centering
    \begin{tabular}{|c|c|} \hline 
         Negligible internal photon losses&$\kappa\approx \kappa_\text{in}$    \\ \hline 
        Resonator-DQD resonance &  $\Omega=\omega_\text{r}=\omega_\text{l}$\\ \hline 
         Negligible electron decoherence& $\Gamma_\text{L,R} \gg \gamma_\phi,\gamma_-$\\ \hline 
         Small interdot tunneling& $\Omega \gg t_c$\\ \hline 
         Unit cooperativity& $C =  4g^2/(\tilde{\Gamma}\kappa)  =1$\\ \hline 
         Linear drive response& $I/e \propto \dot{N}$\\ \hline
    \end{tabular}
    \caption{Conditions for ideal photodetection when the DQD-resonator system is subjected to a monochromatic drive.}
    \label{tab:ideal}
\end{table}
\renewcommand{\arraystretch}{1}
\section{Wigner-function formalism}\label{wfdd}
Throughout the main part of the paper, we focus on the regime where the applied microwave drive is weak. In this low-drive regime, we can solve Eq.~\eqref{pc} and obtain an expression (see App.~\ref{AA}) for the photocurrent in the form
\begin{equation}\label{pcfilter}
    I(t)/e=\frac{1}{2\pi}\int_0^\infty d\tau d\tau' \,h(\tau,\tau')f(t-\tau)f^*(t-\tau'),
\end{equation}
where $h(\tau,\tau')=h(\tau',\tau)$ is a real function of the detector properties only. The detector thus acts as a quadratic, time- and frequency-dependent filter \cite{Cohen} for the pulse with complex amplitude $f(t)$. We note that the photocurrent in Eq.~\eqref{pcfilter} can also be understood as the low-drive amplitude term of a double, or bi-variate, Volterra series \cite{Rice1973,Crespo2017}, with $h(\tau,\tau')$ being the corresponding Volterra kernel. 

For our purposes, it is convenient to rewrite the photocurrent in the compact and compelling form 
\begin{equation}\label{curwig}
    I(t)/e = \int d\tau d\omega \, W_\text{d}(\tau,\omega) W_\text f(t-\tau,\omega), 
\end{equation}
which is a joint time convolution and frequency integration of the Wigner functions of the incident pulse $W_\text{f}(t,\omega)$ and of the detector $W_\text{d}(t,\omega)$. We note that a similar formulation in terms of Wigner functions is common in signal processing \cite{Cohen}, when evaluating spectrograms of different types of non-stationary signals subjected to e.g. filtering or smoothing. Moreover, the formulation in Eq.~\eqref{curwig} is closely related to the one in Ref.~\cite{Hofer}, where the statistics of measurement outcomes is obtained by integrating a product of the detector Wigner function and a quasi-probability representation of the system state. 

\subsection{Pulse Wigner function}
The time-frequency Wigner, or Wigner-Ville, distribution for pulses with complex amplitudes is well known from e.g. signal processing \cite{Bouachache,Boashash,Boashash2,Debnath,Dragoman} and formally identical to the original, quantum-mechanical quasi-probability distribution \cite{Belloni} expressed in terms of the wavefunction. The pulse Wigner function is defined, in terms of the amplitude $f(t)$, as  
\begin{equation}\label{wf}
    W_\text{f}(t,\omega) = \frac{1}{2\pi} \int dy \, e^{-iy\omega} f(t+y/2)f^*(t-y/2).
\end{equation}
We note that, borrowing notation from input-output formalism \cite{Gardiner1985}, we may write $f(t+y/2)f^*(t-y/2)=\langle b^\dagger_{\rm in}(t-y/2)b_{\rm in}(t+y/2)\rangle$, where $b_{\rm in}(t)$ is an annihilation operator that describes the microwave drive. Equation \eqref{wf} is then equivalent to the Wigner function introduced in Ref.~\cite{Ferraro2013} to describe single-electron coherence.

The marginals of the Wigner function in time and frequency are
\begin{equation}
\begin{split}
    W_\text f (t)&=\int  d\omega \, W_\text f (t,\omega) = \abs{f(t)}^2, \\ 
    W_\text f (\omega)&=\int  d t \, W_\text f (t,\omega) = \abs{\tilde f(\omega)}^2.
\end{split}
\label{marginals}
\end{equation}
where $\tilde f(\omega)=1/\sqrt{2\pi}\int dt \, e^{i\omega t} f(t)$ gives the Fourier transform of $f(t)$. The Wigner function is normalized as 
\begin{equation}\label{wfn}
\begin{split}
 \int dt \, d\omega \,  W_\text{f}(t,\omega)=\int dt  W_\text f(t)=\int d\omega  W_\text f (\omega) \\
 =\int dt |f(t)|^2=\int d\omega \abs{\tilde f(\omega)}^2=\bar{N} 
\end{split}
\end{equation}
where $\bar N$ is the average number of photons in the pulse. 

Throughout the paper, we will mainly focus on two common and analytically tractable types of drive pulses, Gaussian and Lorentzian pulses. For a Gaussian pulse, the time-dependent amplitude is
\begin{equation}\label{gauss}
    f(t) = \sqrt{\frac{\bar{N}}{\sigma \pi^{1/2}}}e^{-\frac{t^2}{2\sigma^2}},
\end{equation}
where $\sigma$ denotes the width of the pulse. The Fourier transform of the pulse is given by 
\begin{equation}\label{gaussf}
    \tilde{f}(\omega)=\sqrt{\frac{\bar{N}\sigma}{\pi^{1/2}}}e^{-\frac{\omega^2\sigma^2}{2}}.
\end{equation}
The Wigner function for the Gaussian pulse is 
\begin{equation}\label{wfg}
    W_\text f (t,\omega)=\frac{\bar{N}}{\pi}e^{-t^2/\sigma^2-\omega^2\sigma^2}.
\end{equation}

For a Lorentzian, the pulse amplitude is given by
\begin{equation}\label{lorentz}
    f(t)=\frac{\sqrt{\bar{N}\sigma}}{\sqrt{\pi}(t-i\sigma )}.
\end{equation}
Its Fourier transform is 
\begin{equation}\label{lorentzf}
    \tilde f (\omega)=\sqrt{2\bar{N}\sigma }ie^{- \sigma \omega}\Theta(\omega),
\end{equation}
where $\Theta(\omega)$ is the Heavyside step function.
The corresponding Wigner function is
\begin{equation}\label{wfl}
    W_\text f (t,\omega)=\frac{2\bar N  \sigma e^{-2 \sigma\omega}\sin(2 t \omega)\Theta(\omega)}{\pi t}.
\end{equation}

\begin{figure}[h]
    \centering
    \includegraphics[width=0.9\linewidth]{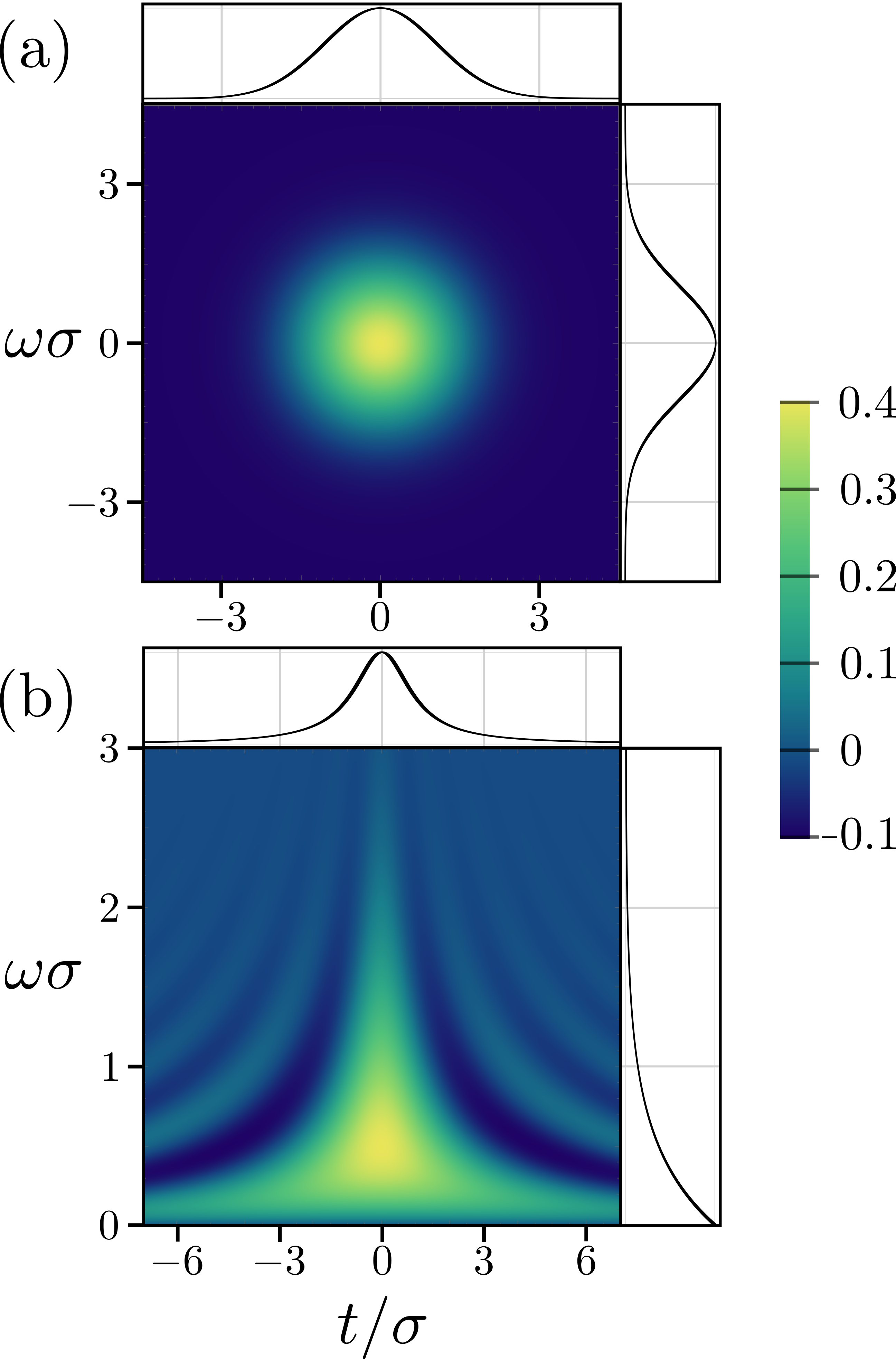}
    \caption{Normalized Wigner functions and their respective time and frequency marginals for (a), a Gaussian pulse and (b), a Lorentzian pulse.}
    \label{fig:wg}
\end{figure}

In Fig.~\ref{fig:wg}, plots of the normalized Wigner functions $W_\text f (t,\omega)/\bar N$ for a Gaussian and Lorentzian pulse are presented, together with their respective time and frequency marginals. 

We point out that our Wigner-function formulation can also be applied to a monochromatic drive as well as a pulse localized in time. For a monochromatic drive at frequency $\omega_\text l$, we have, in the lab frame,
\begin{equation}\label{mcc}
    f(t)=\sqrt{\dot{N}}e^{i\omega_\text l t},
\end{equation}
where $f(t)$ is flux normalized, i.e., describing a constant rate $\dot{N}$ of incident photons. Plugging Eq.~\eqref{mcc} into Eq.~\eqref{wf} gives
\begin{equation}\label{wfm}
    W_\text f(t,\omega)=\dot{N}\delta(\omega-\omega_\text l).
\end{equation}
Here the flux normalization leads to that the average value $\bar W_\text f \rightarrow \infty$. 

For a delta-function pulse, localized in time and arriving at the detector at time $t_0$, we have
\begin{equation}\label{delta}
    f(t)=\sqrt{2\pi N_\nu}\delta(t-t_0),
\end{equation}
where $N_\nu$ denotes the photon number per unit bandwidth.
In this case, the Wigner function is given by
\begin{equation}\label{wfd}
    W_\text f(t,\omega)=N_\nu\delta(t-t_0).
\end{equation}
Again, the average value $\bar W_\text f \rightarrow \infty$. We note that Eqs.~\eqref{wfm} and \eqref{wfd} are equivalent to the Wigner functions of momentum and position eigenstates, respectively.

\subsection{Detector Wigner function}
The detector Wigner function is defined in terms of the kernel $h(\tau,\tau')$ in Eq.~\eqref{pcfilter} as 
\begin{equation}\label{wd}
    W_\text{d}(t,\omega)=\frac{1}{2 \pi} \int dy e^{-i y \omega}h(t+y/2,t-y/2).
\end{equation}
We point out that, in general, $h(\tau,\tau')\neq h_1(\tau)h^*_1(\tau')$, that is, the kernel cannot be written in the form of the pulse Wigner function given in Eq.~\eqref{wf}. This makes the detector Wigner function formally similar to a Wigner function of a mixed quantum state while the pulse Wigner function is equivalent to the Wigner function of a pure state. Moreover, the symmetry property $h(\tau,\tau')=h(\tau',\tau)$ guarantees that $W_\text{d}(t,\omega)$ is real and causality of the detector requires that $W_\text{d}(t,\omega)=0$ for $t<0$. 

The frequency and time marginals of $ W_\text{d}(t,\omega)$, defined analogously to Eq.~\eqref{marginals}, are directly related to the photocurrents obtained in the limits of a monochromatic drive and a delta-function pulse, respectively. Explicitly, inserting the Wigner function for a monochromatic drive at $\omega_\text l$, Eq.~\eqref{wfm}, into the photocurrent expression in Eq.~\eqref{curwig}, the frequency marginal can be written as
\begin{equation}
    W_\text d (\omega_\text l)=\frac{I_\text{mc}(\omega_\text l)/e}{\dot N}= \eta_{\text{mc}}(\omega_\text l),
    \label{Wdfreqmarg}
\end{equation}
This shows that the marginal $W_\text d (\omega_\text l)$ is equal to the detector efficiency $0 \leq \eta_{\text mc} \leq 1$. 

For the time marginal, inserting the Wigner function for a delta-function pulse arriving at the detector at $t_0$, Eq.~\eqref{wfd}, into Eq.~\eqref{curwig}, the time marginal can be written as
\begin{equation}
    W_\text d (t)=\frac{I_\delta(t+t_0)/e}{N_\nu},
    \label{Wdtimemarg}
\end{equation}
where $I_\delta(t+t_0)$ is the photocurrent delta-function, or impulse, response. In addition, from Eq.~\eqref{Wdfreqmarg} we have that the detector Wigner function is normalized as
\begin{equation}\label{norm}
    \bar{W}_\text d =\int dt d\omega \, W_\text d (t,\omega) = \frac{1}{N_\nu}\int dt \, I_\delta(t+t_0)/e,
\end{equation}
which is the number of electrons transferred through the DQD in response to a delta-function pulse, divided by the number of photons per unit bandwidth in the pulse.
We note that the normalization constant $\bar{W}_\text d$ has the unit of frequency.

\subsubsection{Time-frequency uncertainty relation}
As dictated by fundamental principles of quantum mechanics, it is impossible to simultaneously determine the frequency of an incident photon and the photon-arrival time at the detector.

According to Eq.~\eqref{pcfilter}, on the level of averages, photo-detection can be understood as a classical signal processing problem, where a complex signal $f(t)$ is being measured by the photocurrent $I(t)$. It is, however, still impossible to simultaneously determine the frequency as well as the arrival time of an incoming pulse due to the Heisenberg-Gabor limit \cite{Gabor1946}, which states that one cannot sharply localize a signal in both the time domain and the frequency domain. To detect the arrival time of a pulse perfectly, the detector Wigner function needs to obey $W_{\rm d}(t,\omega) \propto \delta (t)$. In this case, the response to a delta-function pulse [c.f.~Eq.~\eqref{Wdtimemarg}] is again a delta-function pulse. Alternatively, the detector becomes frequency selective if $W_{\rm d}(t,\omega) \propto \delta (\omega-\omega_{\rm l})$. In this case, only signals with frequency $\omega_{\rm l}$ are detected. To be both frequency selective and to detect arrival times perfectly, the Wigner function would thus need to obey $W_{\rm d}\propto \delta(t)\delta(w-\omega_{\rm l})$. This is prevented by the Heisenberg-Gabor limit which reads, in full analogy to the Heisenberg uncertainty relation,
\begin{equation}
   \Delta t ~\Delta \omega \geq \frac{1}{2},
   \label{uncertain}
\end{equation}
where $\Delta \omega=\sqrt{\langle \omega^2 \rangle-\langle \omega \rangle^2}$, $\Delta t=\sqrt{\langle t^2 \rangle-\langle t\rangle^2}$ and $\langle x \rangle=\int dt  d\omega \, x W_\text d(t,\omega)/ \bar W_\text d$. A detailed derivation of Eq.~\eqref{uncertain} is given in App. ~\ref{AD}. We note that Eq.~\eqref{uncertain} only holds for detectors that do not allow for a negative photocurrent. Should the photocurrent become negative, which is in principle possible in our system due to backtunneling from the drain, $\Delta t$ is no longer a good measure of the spread of the photocurrent.

In the following, we consider a good detector to have a fast time-resolution, i.e., a small $\Delta t$, as this allows one to determine the shape of an incoming microwave pulse. In this case, the ideal detector Wigner function reads
\begin{equation}
    \label{eq:idealwd}
    W_{\rm d} = \delta(t)\hspace{.5cm}\Rightarrow\hspace{.5cm}I(t)/e = |f(t)|^2.
\end{equation}
We note from Eq.~\eqref{Wdfreqmarg} that such an ideal detector has unit efficiency at all frequencies, $\eta_{\rm mc}(\omega)=1$ and thus, an infinitely wide bandwidth, as well as a diverging normalization $\bar{W_\text d}$. For real detectors, $\eta_{\rm mc}(\omega)\rightarrow 0$ for $|\omega|\rightarrow \infty$ and the Wigner function cannot develop diverging peaks as it obeys the inequality (see App.~\ref{AD} for a derivation)
\begin{equation}
    \label{eq:wignerbound}
    |W_{\rm d}(t,\omega)|\leq \frac{\bar{W}_{\rm d}}{\pi}.
\end{equation}
As for Eq.~\eqref{uncertain}, this relation only holds for detectors that do not allow for a negative photocurrent.

\subsection{Our detector}
Turning to the photodetector described in the previous section, we focus our discussion on the parameter regime described in Tab.~\ref{tab:ideal}, leading to ideal photodetection for a monochromatic drive, with the slight difference that we relax the requirement for unit cooperativity, i.e. $C \neq 1$. In this parameter regime, we obtain an analytical, closed form expression for the detector Wigner function (for $t\geq0$, as we find $W_\text{d}(t,\omega)=0$ for $t<0$ due to causality) 

\begin{widetext}
\begin{equation}\label{idealwigner}
 W_\text{d}(t,\omega)=32\pi g^2\Gamma_\text R\kappa e^{-t(\Gamma_\text R+\kappa)/2}\left\{  \begin{array}{ll}  \dfrac{\alpha\cos(\alpha t/2)\sin(2 t \omega)-4\omega\sin(\alpha t/2)\cos(2 t \omega)}{\alpha\omega(16\omega^2-\alpha^2)}, & 4g > \abs{\Gamma_\text R-\kappa} \\[0.35cm] 
\dfrac{\alpha'\cosh(\alpha' t/2)\sin(2 t \omega)-4\omega\sinh(\alpha' t/2)\cos(2 t \omega)}{\alpha'\omega(16\omega^2+\alpha'^2)}, &   4g < \abs{\Gamma_\text R-\kappa} \end{array} \right.
\end{equation}
where $\alpha = \sqrt{-(\Gamma_\text R-\kappa)^2+16g^2}$, $\alpha' = \sqrt{(\Gamma_\text R-\kappa)^2-16g^2}$. Details on the derivation of Eq.~\eqref{idealwigner} are found in App.~\ref{AA}. We observe that the Wigner function is symmetric in frequency, $W_\text d(t,\omega)=W_\text d(t,-\omega)$ and is invariant under the exchange $\Gamma_\text R \leftrightarrow \kappa$. This symmetry can be understood physically. For weak drives, the DQD behaves just like a harmonic oscillator, see App.~\ref{AD2} and Ref.~\cite{Zenelaj}. Our photodetector can then be mapped to two coupled cavities, one of which is coherently driven and has linewidth $\kappa$ and one which is not driven with linewidth $\Gamma_\text R$. In this picture, the photocurrent $I(t)$ corresponds to the photon flux out of the undriven cavity and the efficiency is determined by the transmission through the two cavities. The symmetry under the exchange $\Gamma_\text R \leftrightarrow \kappa$ then corresponds to the reciprocity of two coupled cavities: The transmission through the two cavities is invariant under exchanging the two cavities, or equivalently, under exchange of source and detector \cite{deak_2012}. This is true in the absence of dephasing and relaxation $\gamma_\phi=\gamma_-=0$ and thus in the regime considered here.

Integrating Eq.~\eqref{idealwigner} over time (frequency) we have the frequency (time) marginal of the Wigner function as

\begin{equation}
     W _\text d (t) =64 \pi^2 g^2 \Gamma_\text R\kappa e^{-t(\Gamma_\text R+\kappa)/2 } \left\{  \begin{array}{ll}  \dfrac{\sin^2(\alpha t/4)}{\alpha^2}, & 4g > \abs{\Gamma_\text R-\kappa}  \\[0.35cm] 
\dfrac{\sinh^2(\alpha' t/4)}{\alpha'^2}, &   4g < \abs{\Gamma_\text R-\kappa}\end{array} \right.
\label{Wdtime}
\end{equation}
and
\begin{equation}
     W _\text d (\omega) =512\pi g^2 \Gamma_\text R \kappa \left\{  \begin{array}{ll}  \dfrac{1}{[(\Gamma_\text R+\kappa)^2+(\alpha-4\omega)^2][(\Gamma_\text R+\kappa)^2+(\alpha+4\omega)^2]}, & 4g > \abs{\Gamma_\text R-\kappa}  \\[0.35cm] 
\dfrac{1}{[(\alpha'+\Gamma_\text R+\kappa)^2+16\omega^2][(-\alpha'+\Gamma_\text R+\kappa)^2+16\omega^2]}, &   4g < \abs{\Gamma_\text R-\kappa}\end{array} \right.
\label{Wdfreq}
\end{equation}
\end{widetext}
Via Eqs.~\eqref{Wdfreqmarg} and \eqref{Wdtimemarg}, Eqs.~\eqref{Wdfreq} and \eqref{Wdtime} thus give the explicit expressions for the photocurrent of a monochromatic drive and the photodetector impulse response, respectively. 

\begin{figure*}[ht!]
\centering
    \includegraphics[width=0.99\linewidth]{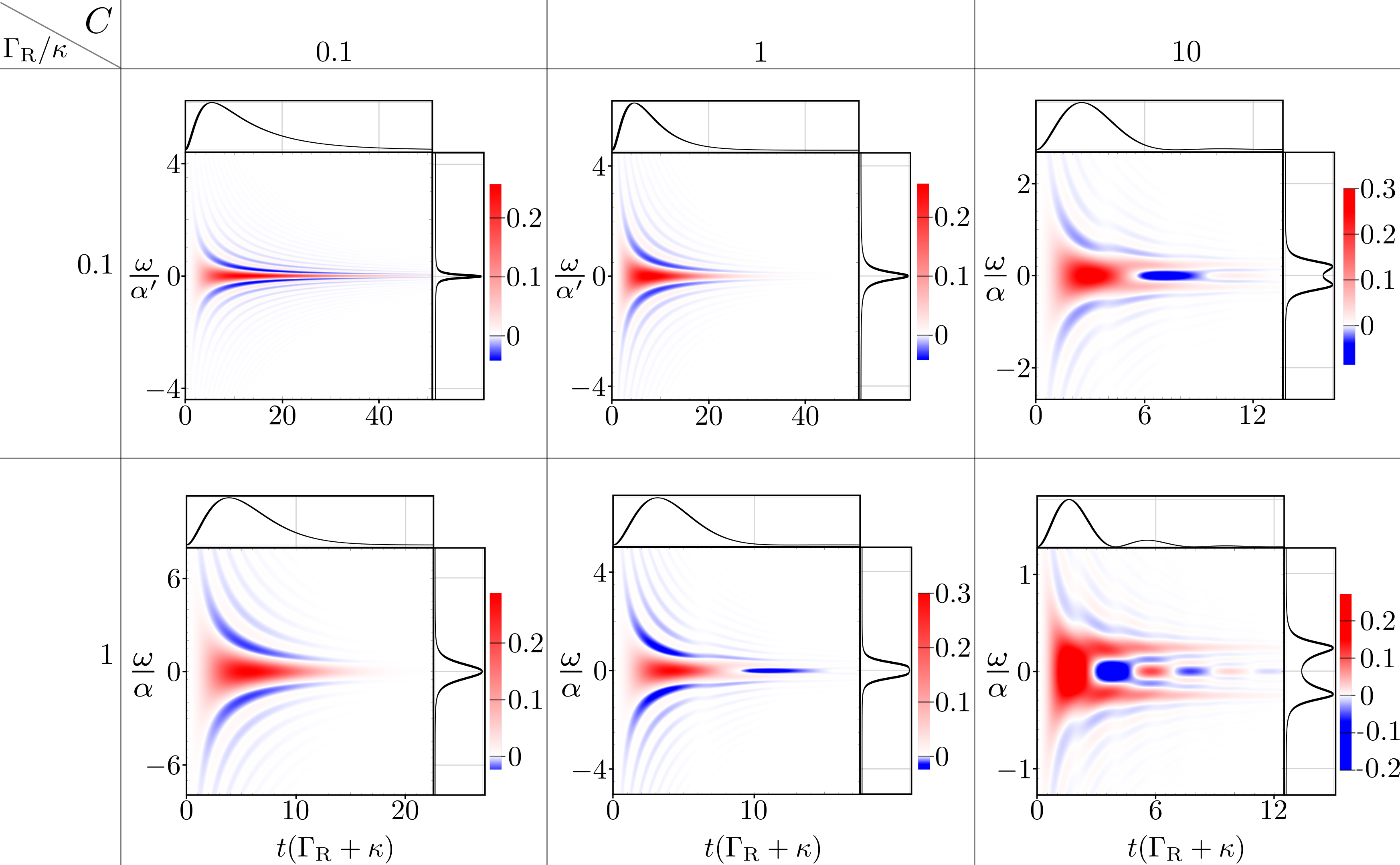}
    \caption{Normalized Wigner function of the detector with its respective time and frequency marginals for $\Gamma_{\rm R}/\kappa = 0.1$, $1$ and cooperativity $C = 0.1,$ $1,$ and $10$.}
    \label{fig:1}
\end{figure*}

The normalization of the Wigner function is given by
\begin{equation}
    \bar W_\text d =\frac{16 \pi^2 g^2 \Gamma_\text R \kappa}{(\Gamma_\text R+\kappa)(\Gamma_\text R\kappa+4 g^2)}.
\end{equation}
In Fig.~\ref{fig:1}, we plot the normalized Wigner functions in Eq.~\eqref{idealwigner} together with their respective, normalized time and frequency marginals for different values of $\Gamma_{\rm R}/\kappa$ and $C$. 

We see that $W_\text d$, for all parameter values, has a T-shape in time-frequency space with a global maximum at $\omega=0$ and $t>1/(\Gamma_\text R+\kappa)$, located at $t=6/(\Gamma_\text R+\kappa)$ at $\alpha\rightarrow 0$. The Wigner function also displays distinct oscillations around zero in time and frequency due to the harmonic terms $\sin(2\omega t)$ and $\cos(2\omega t)$. The oscillations decay towards large frequencies and times. While $W_\text d$ has a complex structure in both time and frequency space, we note that the exponential decay with increasing time is governed by the time scale $1/(\Gamma+\kappa)$. As a result, the minimum effective bandwidth of the detector is of order $\Gamma_\text R+\kappa$, which is also clear from the expression for the frequency marginal in Eq.~\eqref{Wdfreq}. We also observe a splitting of the resonance peak in the frequency marginal, which is on the order of $\alpha/2$ and is most clearly visible when $C>1$. This appears to be a consequence of the Rabi splitting, which is present in the regime of strong electron-photon coupling \cite{toida}.

\section{Photocurrent response and detector operation}\label{prdo}
To both qualitatively and quantitatively characterize the operation of our system as a detector for incoming photon pulses, we compare the outgoing electrical photocurrent $I(t)$ to the incoming pulse magnitude $|f(t)|^2$. We work in the ideal-detector limit, implying that we use $C=1$ throughout this section. In addition to the monochromatic drive, where the detector efficiency is the key performance measure, for a time-dependent photon pulse also the ability of the detector to respond rapidly and to faithfully transfer the temporal shape of the drive into the output photocurrent are important features. In order to quantify the performance of the photodetector, we will thus employ three different performance measures:
\begin{enumerate}
\item The photodetection efficiency, which in the case of a pulse input is given by 
\begin{equation}\label{effpulse}
\eta_\text p = \frac{Q/e}{\bar{N}},
\end{equation}
where $Q=\int dt \, I(t)$ is the average transferred charge in the electrical photocurrent pulse. The efficiency $0 \leq \eta_\text p \leq 1$ thus provides information on the ability of the photodetector to detect the average number of photons in the drive pulse, key for number resolved photodetection. Unit efficiency implies that every photon in the pulse is converted into exactly one electron in the photocurrent.
\item The average delay in time between the input and output signal, which is defined as
\begin{equation}\label{tdelay}
   \tau_\text d = \int dt \, t \left[ \frac{I(t)/e}{Q} - \frac{\abs{f(t)}^2}{\bar N} \right].
\end{equation}
We note that the input and output signals have been normalized to correctly produce the average times. 
\item The fidelity of the temporal shape transfer, given by 
\begin{equation}\label{overlap}
O = \int dt \, \sqrt{\frac{\abs{f(t-\tau_\text{d})}^2}{\bar N}\cdot\frac{I(t)/e}{Q}}.
\end{equation}

\begin{figure*}[ht!]
    \centering
    \includegraphics[width=0.99\linewidth]{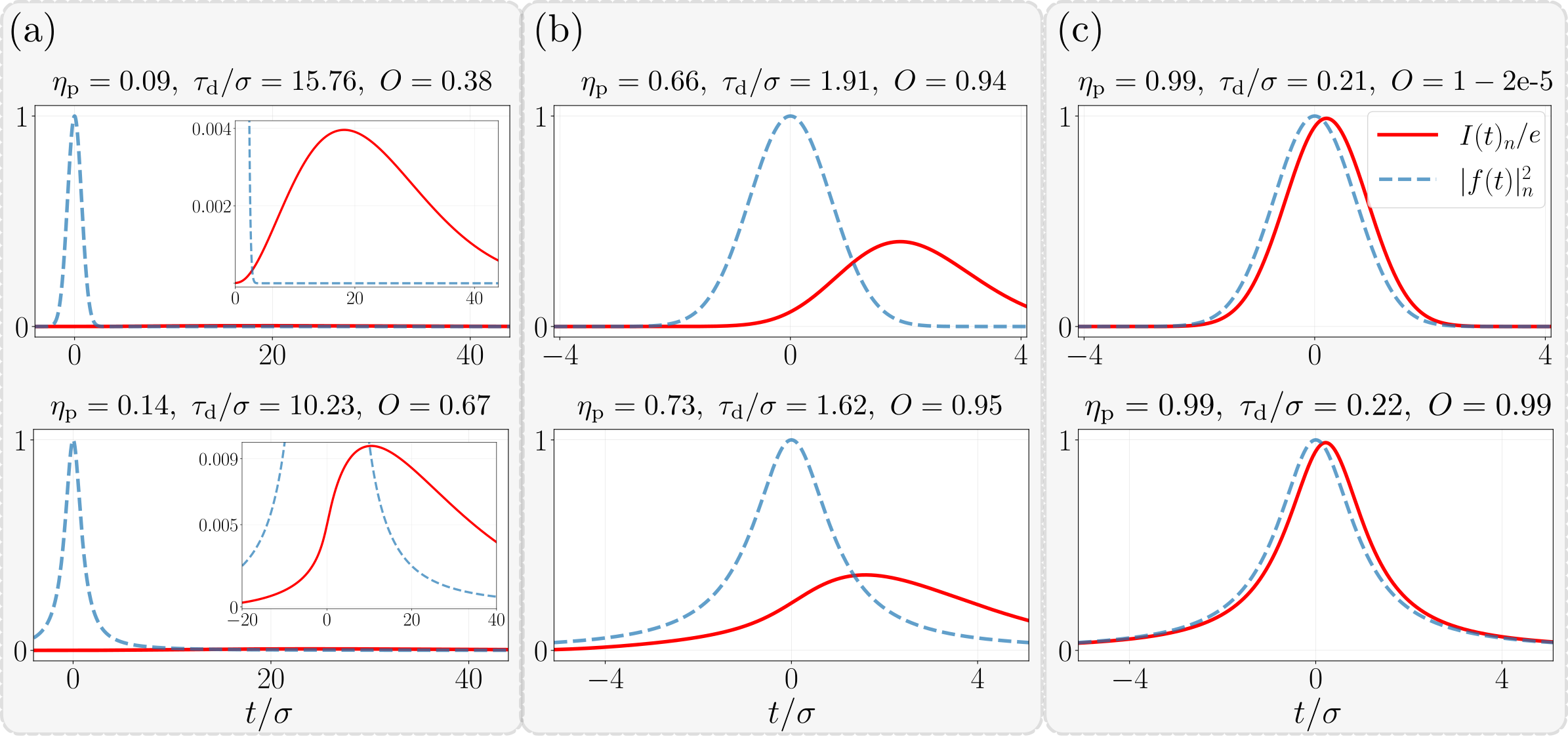}
    \caption{Normalized photocurrent in the ideal-detector limit and drive pulses for a Gaussian (top) and Lorentzian (bottom) drive for $\bar{N}=0.1$, $\Gamma_\text R/\kappa=1$ and (a) $\sigma (\Gamma_\text R+\kappa)/2=0.1$, (b) $\sigma (\Gamma_\text R+\kappa)/2=1$, (c) $\sigma (\Gamma_\text R+\kappa)/2=10$. Insets in (a) show a zoomed-in version to visualize the electrical photocurrent pulse.}
    \label{fig:c_g}
\end{figure*}
Here, the input and output signals have been normalized and shifted with the average time delay, in order to compare the shapes only. Moreover, we employ the Bhattacharyya distance \cite{Bhatta}, quantifying the similarity, or overlap, of two distributions. This gives $0 \leq O \leq 1$, with $O=1$ for identical input and output shapes. 
\end{enumerate}

We note that for the idealized detector described by the Wigner function in Eq.~\eqref{eq:idealwd}, we find $\eta_{\rm p}=1$, $\tau_{\rm d}=0$, and $O=1$, implying optimal performance according to our figures of merit.

We expect the relation between the spectral width of the drive pulse and the typical bandwidth of the detector, $\sim \Gamma_\text R+\kappa$, to be the defining aspect of the detector performance. We will therefore focus on three different regimes, where the frequency width of the incoming pulse is larger, comparable to, and smaller than the detector bandwidth, respectively. In Fig.~\ref{fig:c_g}, we plot the photocurrent in the ideal-detector limit together with the incoming pulse in the three different regimes for a Gaussian and a Lorentzian drive, respectively. The photocurrent and drive pulses have been normalized with the same pre-factor, such that $\abs{f(t=0)}_n^2=1$, where the subscript stands for \textit{normalized}. We also give the numerical values for the different performance quantifiers. 

We observe a similar trend for both Gaussian and Lorentzian drive pulses. The detector's performance significantly improves from panels (a) to (c) across all quantifiers. This improvement is anticipated, as the detector Wigner function approaches the ideal limit described in Eq.~\eqref{eq:idealwd} with increasing detector bandwidth. We also observe that the conditions for ideal photodetection for a monochromatic pulse given in Tab.~\ref{tab:ideal} result in very good detection of microwave pulses according to all three figures of merit used. 

\subsection{Experimental setting}

In an experimental setting, the parameters governing the DQD-resonator system are typically not ideal. We would thus like to study the photodetection of single pulses in a more realistic parameter regime. To that end, we use the parameters obtained from the experiment in Ref.~\cite{haldar3}, where experimentally, a photo-detection efficiency of 25\% was achieved using a monochromatic drive. The numerical values for the system parameters are summarized in Tab.~\ref{tab:exp}. 

\begin{figure}[ht!]
    \centering
    \includegraphics[width=0.9\linewidth]{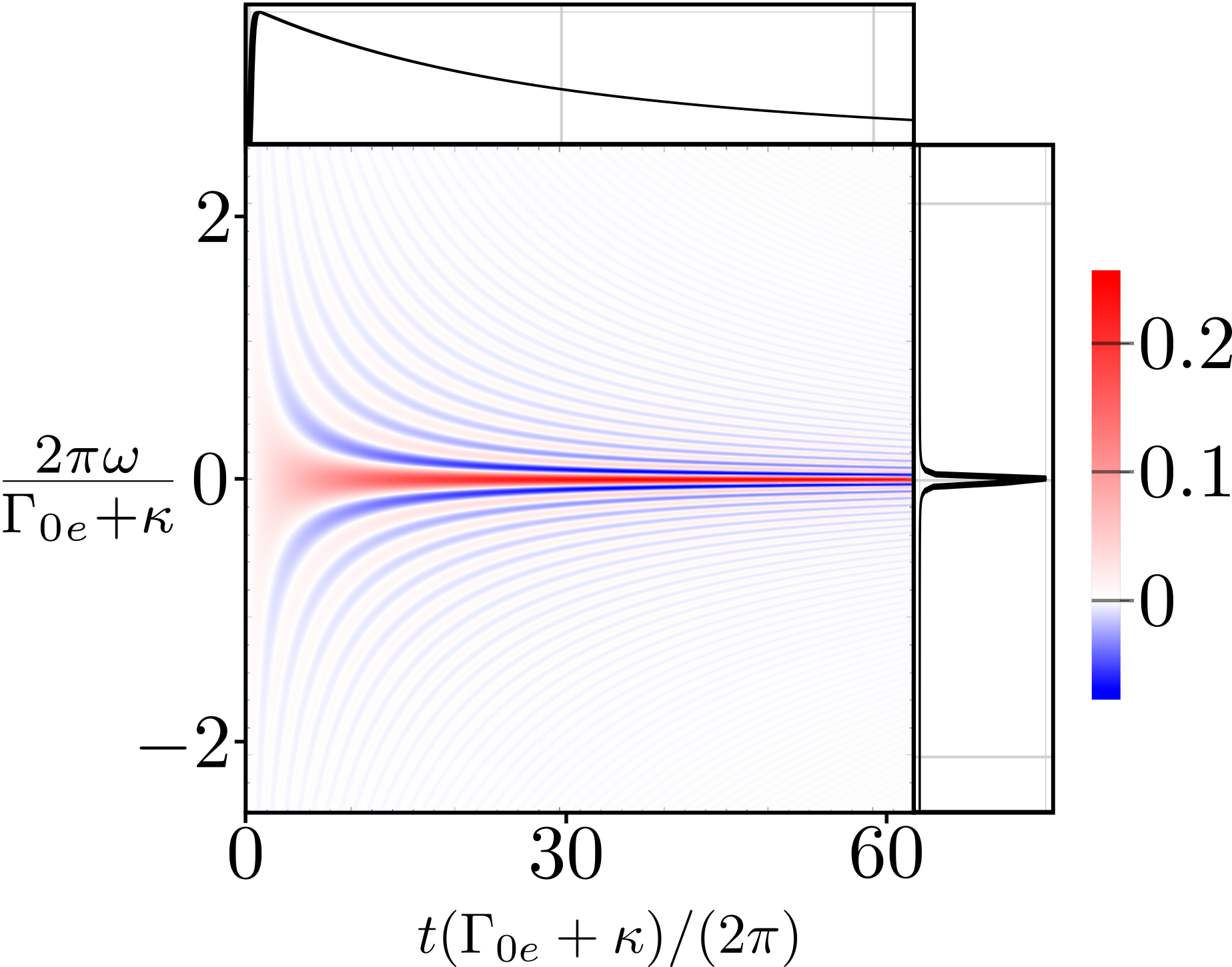}
    \caption{Normalized Wigner function of the detector and its respective time and frequency marginals using the experimental parameters in Tab.~\ref{tab:exp}.}
    \label{fig:wig_exp}
\end{figure}

\begin{table}[h]
    \centering
    \begin{tabular}{|>{\centering\arraybackslash}p{0.2\linewidth}|>{\centering\arraybackslash}p{0.3\linewidth}|} \hline 
         $\kappa_\text{in}/2 \pi$& 3.8 MHz\\ \hline 
         $\kappa/2\pi$& 5.2 MHz\\ \hline 
         $\Gamma_\text{L}/2\pi$& \hspace{0.01mm} 12 MHz\\ \hline 
         $\Gamma_\text{R}/2\pi$& 1.1 GHz\\ \hline 
         $g/2\pi$& \hspace{0.01mm} 43 MHz\\ \hline 
         $\gamma_-/2\pi$& \hspace{0.01mm} 23 MHz\\ \hline 
         $\gamma_\phi / 2\pi$& 1.8 GHz\\ \hline
         $\Omega/2 \pi$& \hspace{-0.8em} 6.67 GHz\\\hline
         $\epsilon/2 \pi $&\hspace{-0.8em}$-$6.6 GHz\\\hline
         $C$ & 0.15 \\ \hline
    \end{tabular}
    \caption{Experimental rates governing the DQD-resonator system, taken from Ref.~\cite{haldar3}.}
    \label{tab:exp}
\end{table}
We again look at the Wigner function of the detector by numerically solving Eq.~\eqref{wd}, see Fig.~\ref{fig:wig_exp}.
The Wigner function looks similar to the one in the upper left panel of Fig.~\ref{fig:1}. This is to be expected, as we are in a similar parameter regime.

\begin{figure}[ht!]
    \centering
    \includegraphics[width=0.8\linewidth]{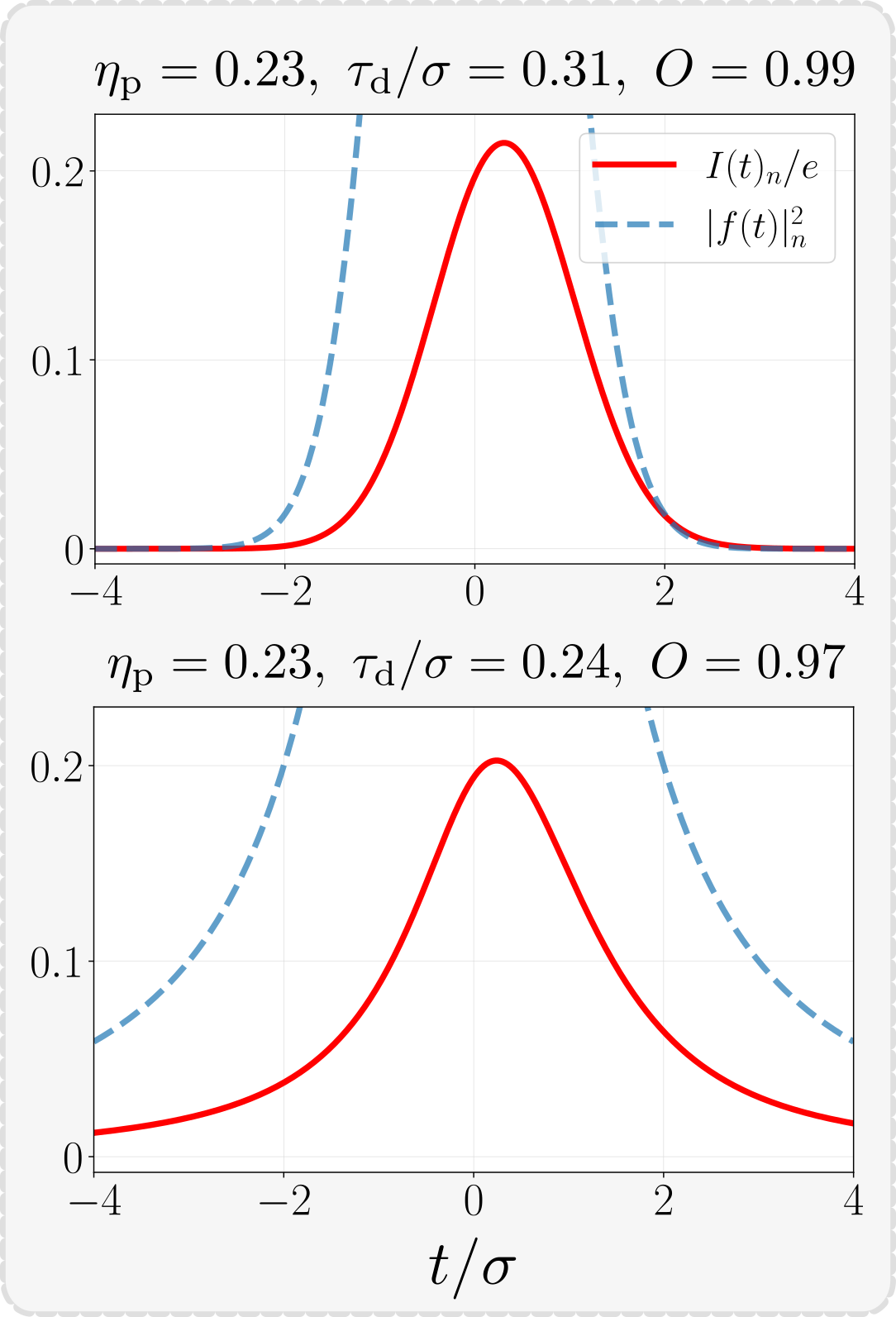}
    \caption{Normalized photocurrent pulses for a  Gaussian (top) and Lorentzian (bottom) drive for $\bar N = 0.1$ and $\sigma(\Gamma_{0e} + \kappa)/(4\pi)=77$ using the experimental parameters in Tab.~\ref{tab:exp}.}
    \label{fig:exp}
\end{figure}

In Fig.~\ref{fig:exp}, we plot the photocurrent together with the incoming pulse for a Gaussian and a Lorentzian drive, respectively and give the numerical values for the different performance quantifiers. 

\section{Beyond the low-drive limit}\label{bld}
For larger microwave drive amplitudes, the low-drive expression for the photocurrent in Eq.~\eqref{pcfilter} as well as the Wigner function formulation in Eq.~\eqref{curwig} do not hold anymore. To investigate the time-dependent photocurrent response we instead solve the Lindblad master equation in Eq.~\eqref{LME} numerically using QuTip \cite{qutip} and evaluate the photocurrent via Eq.~\eqref{pc}.

\begin{figure*}[ht!]
    \centering
    \includegraphics[width=0.99\linewidth]{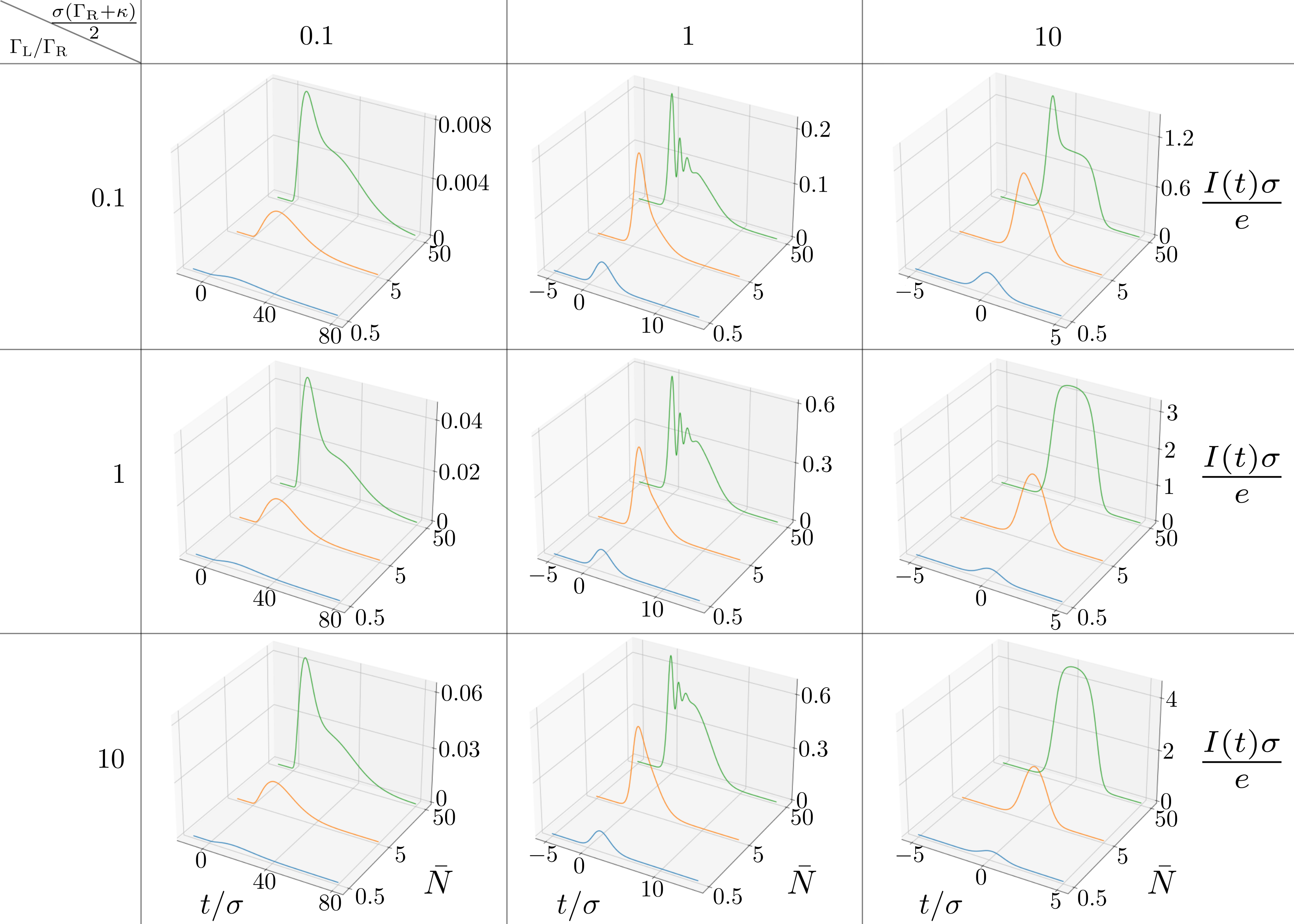}
    \caption{Normalized photocurrent as a function of time and average number of photons in the Gaussian drive for $\sigma(\Gamma_{\text{R}} +\kappa)/2 = 0.1,$ 1, and 10, $\Gamma_{\text{L}} /\Gamma_{\text{R}}=0.1,$ 1, and 10, and $\bar{N} = 0.5$, 5, and 50.}
    \label{figdiffN}
\end{figure*}

To obtain a qualitative picture of the photocurrent response beyond the low-drive regime, we plot in Fig.~\ref{figdiffN} the photocurrent as a function of time for drive strengths and a set of detector parameters $\Gamma_\text L, \Gamma_\text R, \kappa$. For simplicity, we consider an incoming Gaussian pulse of temporal width $\sigma$, however, qualitatively similar results are obtained for a Lorentzian pulse as also discussed further below.
The overall pattern is that the photocurrent pulses increase in amplitude and change shape when increasing the drive amplitude. There are two distinct features, different from the low-drive limit, to be emphasized:

First, the photocurrent becomes dependent on the rate, $\Gamma_\text L$, for electrons to tunnel into the DQD. This is most clearly visible for $\sigma(\Gamma_\text R+\kappa) \gtrsim 1$ in Fig.~\ref{figdiffN} and can be understood as follows: In the low-drive limit, the photon assisted electron tunneling between the left and the right dot is the slowest transport process, acting as a bottle-neck for the photocurrent. Hence, the dot filling rate $\Gamma_\text L$ does not play any role for the photocurrent, as also evident from Eq.~\eqref{idealwigner}. Increasing the drive strength, the photon assisted tunnel rate becomes comparable to the other rates in the problem and $\Gamma_\text L$ starts to affect the photocurrent $I(t)$. 

Second, for certain parameters, most prominent for $\sigma(\Gamma_\text{R}+\kappa) \approx 1$ in Fig.~\ref{figdiffN}, the photocurrent develops temporal oscillations for increasing drive amplitudes. The amplitude and frequency of the oscillations depend on both drive strength $\bar{N}$ as well as the ratio $\Gamma_\text L/\Gamma_\text R$. Importantly, the oscillations persist for large-drive amplitudes $\bar{N} \gg 1$. 

\subsection{Large-drive limit}
While the $\Gamma_\text L$ dependence of the photocurrent is explained above, the photocurrent oscillations require additional analysis. To that aim, we focus on the large-drive limit, $\bar{N} \gg 1$, where the influence of the DQD on the resonator field can be neglected and the photon operators $a,a^{\dagger}$ can be taken equal to their average values, $\langle a \rangle, \langle a^{\dagger} \rangle$. The time evolution of the average $\langle a \rangle(t)$ is given by the equation
\begin{equation}
    \partial_t \langle a \rangle (t) =-\frac{\kappa}{2}\langle a\rangle (t)-i \sqrt{\kappa}f(t),
\label{amp}
\end{equation}
where we recall that $f(t)$ is the time-dependent drive amplitude. Following our earlier work, \cite{Zenelaj}, we can then write the equations of motion for the DQD density matrix components as
\begin{equation} \label{largedriveeq}
    \partial_t p_0(t)=-\Gamma_{\rm L}p_0(t)+\Gamma_{\rm R} p_e(t),
\end{equation}
\begin{equation}
\begin{aligned}
    \partial_t &p_e(t)=-\Gamma_{\rm R} p_e(t) \\
     &-i g \left[\langle a \rangle (t)\expval{\sigma_+}(t)-\langle a^{\dagger}\rangle(t)\expval{\sigma_-}(t)\right], 
\end{aligned}
\end{equation}
\begin{equation}
\begin{aligned}
    \partial_t &p_g(t)=\Gamma_{\rm L} p_0(t) \\ 
    &+i g \left[\langle a \rangle (t)\expval{\sigma_+}(t)-\langle a^{\dagger}\rangle(t)\expval{\sigma_-}(t)\right],
\end{aligned}
\end{equation}
\begin{equation}
    \partial_t \expval{\sigma_-}(t)=-\frac{\Gamma_{\rm R}}{2}\expval{\sigma_-}(t)+ig \langle a \rangle (t)[p_e(t)-p_g(t)],
\end{equation}
where the probabilities $p_\alpha(t)=\langle \alpha |\tilde \rho|\alpha \rangle$, $\alpha=0,e,g$, with $p_e(t)+p_g(t)+ p_0(t)=1$, and $\expval{\sigma_-}(t)=\langle e|\tilde \rho|g \rangle=[\expval{\sigma_+}(t)]^*$. We recall that $\tilde \rho$ is the reduced density matrix of the DQD, obtained by tracing the full density matrix $\rho$ over the degrees of freedom of the resonator. From the density matrix components, the photocurrent can be calculated via Eq.~\eqref{pc}.

\begin{figure}[ht!]
    \centering
   \includegraphics[width=0.99\linewidth]{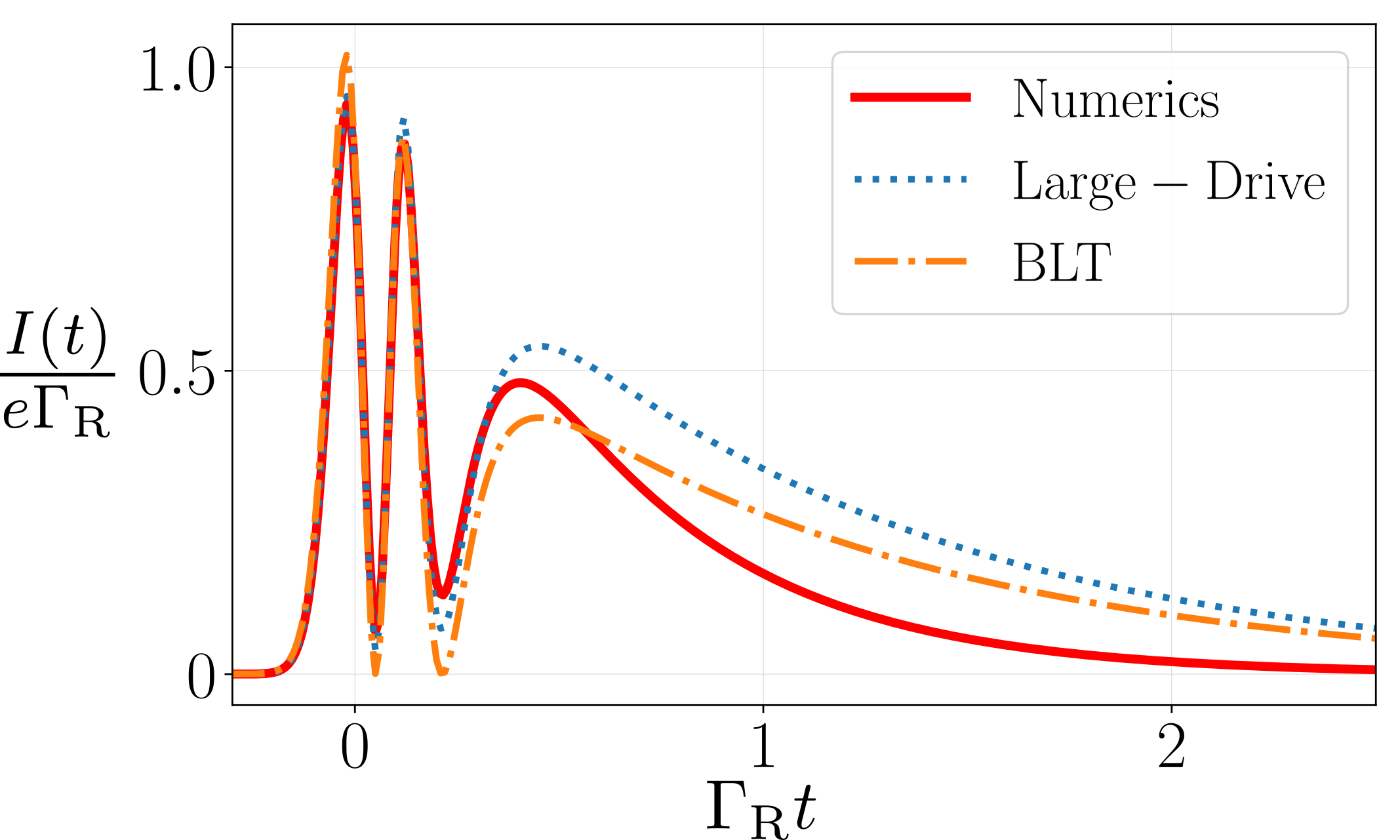}
    \caption{Photocurrent from the full numerics (full line), the large-drive limit (dotted line), and the boundary-layer theory (dashed-dotted line) with $\bar N = 150$, $\sigma \Gamma_{\rm R} = 0.1$, $\kappa/\Gamma_{\rm R}=20$, $\Gamma_{\rm L} /\Gamma_{\rm R}=12$ and $t_c/\Gamma_{\rm R} = \gamma_-/\Gamma_{\rm R} = \gamma_\phi/\Gamma_{\rm R} = 0$.}
    \label{blt}
\end{figure}

To illustrate that these equations correctly describe the photocurrent in the large-drive limit, we compare their solution to the exact numerical solution for a Gaussian pulse, see Fig.~\ref{blt}. 

As is clear from the figure, the large-drive equations capture the time dependence of the photocurrent well, including the $\Gamma_\text L$ dependence and the temporal oscillations.

To obtain a simple picture of the oscillations we further consider the limit where $\Gamma_\text L$ is much larger than $\Gamma_\text R$. In this limit the DQD is rapidly refilled when an electron tunnels out, which implies that the DQD is essentially never empty, $p_0(t)\ll 1$. We can then, to leading order in $\Gamma_{\rm R}/\Gamma_{\rm L}$, perform an adiabatic elimination of $p_0(t)$ from Eq.~\eqref{largedriveeq}, resulting in $p_0(t)=\Gamma_{\rm R}/\Gamma_{\rm L}p_e(t)$. Inserting this back into the remaining equations we get
\begin{equation}\label{Rabi}
\begin{split} 
    \partial_t p_e(t)&=-\Gamma_{\rm R} p_e(t) -i\left[\Omega^*(t)\expval{\sigma_+}(t)-\-\Omega(t)\expval{\sigma_-}(t)\right], \\ 
    \partial_t p_g(t)&=\Gamma_{\rm R} p_e(t)+i\left[\Omega^*(t)\expval{\sigma_+}(t)-\-\Omega(t)\expval{\sigma_-}(t)\right], \\  
    \partial_t \expval{\sigma_-}(t)&=-\frac{\Gamma_{\rm R}}{2}\expval{\sigma_-}(t)+i\Omega^*(t)[p_e(t)-p_g(t)],
\end{split}
\end{equation}
where we introduced $\Omega(t)=g\langle a^\dagger \rangle(t)$. The equations in \eqref{Rabi} are the well known equations for the semiclassical Rabi problem \cite{Griffiths,Merlin}, describing a driven two-level system with a decay rate $\Gamma_{\rm R}$. Here we consider a drive with a time-dependent amplitude, giving rise to a time-dependent Rabi frequency $\Omega(t)$.

The temporal oscillations in the photocurrent, visible in Figs. \ref{figdiffN} and \ref{blt}, can thus be understood as manifestations of the coherent excitation and de-excitation of the DQD, induced by the photon pulse and decaying with time due to the tunneling out of the electron from the DQD. The effect of the time dependence of the Rabi frequency is primarily to modulate the frequency of the photocurrent oscillations.

Further insight can be obtained by considering the regime of weak decay, $\Gamma_{\rm R}\ll \kappa$, analyzing the different temporal regimes. For times much shorter than the decay time, $t\ll 1/\Gamma_{\rm R}$, the photocurrent is given from the so-called area law \cite{Fischer}, as
\begin{equation}
    I(t)/e=\Gamma_{\rm R}\sin^2\left[\mathcal{A}(t)\right], \hspace{0.3cm} \mathcal{A}(t)=\int_{-\infty}^t dt' \Omega(t').
    \label{arealaw}
\end{equation}
The photocurrent thus displays modulated oscillations with a time-dependent frequency. For times much longer than the decay time of photons in the resonator, $t \gg 1/\kappa$, the resonator is empty, $\langle a(t)\rangle=0$, and hence $\Omega(t)=0$. In this regime, it follows directly from Eq.~\eqref{Rabi} that $p_e(t)\sim e^{-t\Gamma_{\rm R}}$, exponentially decaying in time. Based on these two limiting cases, we can obtain an approximate expression for the photocurrent for all times, including the regime $1/\Gamma_{\rm R}<t<1/\kappa$ where none of the limiting expressions are formally valid. To do this, we apply boundary-layer theory \cite{Schlichting}, matching the asymptotics of the two expressions. This gives the photocurrent, see App.~\ref{AE}, as
\begin{equation}
    I(t)/e=\Gamma_{ \rm R}\sin^2\left[\mathcal{A}(t)\right]e^{-t\Gamma_{\rm R}}.
    \label{BLT}
\end{equation}
This expression captures both the temporal oscillations and the exponential decay and approximates well the exact result, as illustrated in Fig.~\ref{blt}. 

In addition, we note that an analytical treatment is amenable for a very short drive pulse, taking $f(t)=A\delta(t)$, a $\delta$-function with strength $A$. This leads, via Eq.~\eqref{amp}, to a Rabi frequency $\Omega (t)=-ig\sqrt{\kappa}Ae^{-t\kappa/2}\Theta(t)$, displaying a sharp onset of amplitude $g\sqrt{\kappa}A$ at $t=0$, followed by an exponential decay with rate $\kappa/2$. Inserting this into the Rabi equations \eqref{Rabi}, the resulting system of differential equations can be solved analytically. The resulting expression for the photocurrent $I(t)/e=\Gamma_{\rm R}p_e(t)$ is however lengthy and provides little qualitative insight and is therefore only presented, together with the derivation, in App.~\ref{AF}.

\subsection{Detection efficiency} 
\begin{figure}[ht!]
    \centering
    \includegraphics[width=0.99\linewidth]{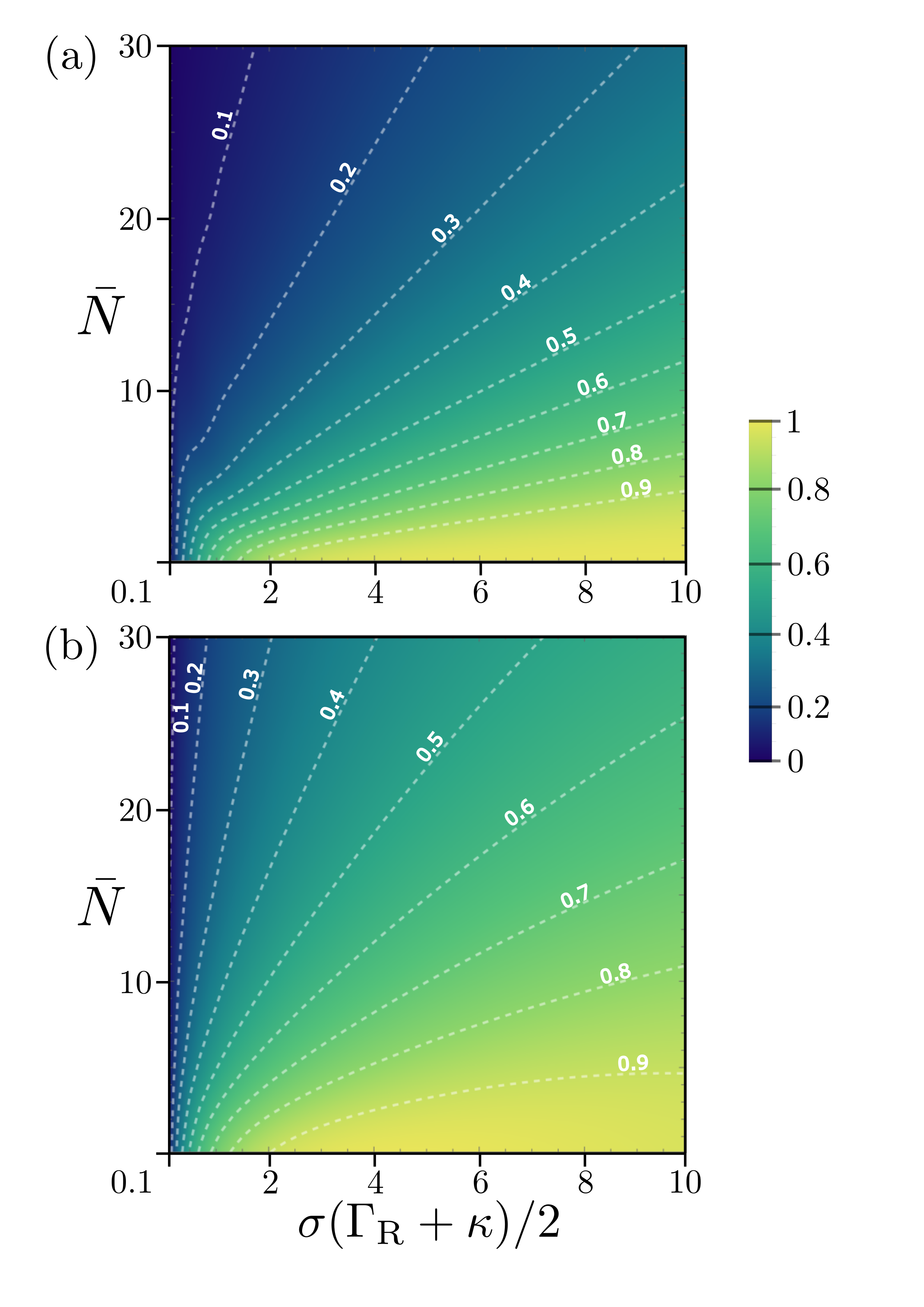}
    \caption{Photodetection efficiency as a function of $\bar N$ and $\sigma(\Gamma_\text R+\kappa)/2$ with $\Gamma_\text R/\kappa = 1$ and $\Gamma_\text L /\Gamma_\text R = 1$, for (a) a Gaussian drive and (b) a Lorentzian drive.}
    \label{fig:e_p}
\end{figure}

Out of the three detector performance quantifiers analyzed in the low-drive limit, the detection efficiency is the most interesting one to investigate beyond low drive - the strongly modified temporal properties of the photocurrent make the delay time and fidelity less relevant. The photodetection efficiency for both a Gaussian and a Lorentzian drive pulse are shown in Fig.~\ref{fig:e_p} for a representative set of system parameters. 
From the figure it is clear that the overall trend is a suppressed efficiency for increasing drive strength. However, for a wide pulse, $\sigma(\Gamma_\text R+\kappa)\gg 1$, the efficiency remains close to unity also for large $\bar N\gg 1$. In this regime the detector successively, and efficiently, converts individual pulse photons into electrons traversing the DQD.

\section{Conclusions and outlook}\label{cao}
To summarize, we have theoretically investigated the detection of single microwave pulses in a resonator-coupled DQD. In the regime of low applied drive, the drive and detector can be described in terms of Wigner functions, which allows for a clear and visually compelling analysis of their time and frequency dependence. 
This Wigner-function description was used to compute the outgoing photocurrent pulse, which was compared to the incoming microwave pulse in different parameter regimes. In the regime of a broadband photodetector, the performance of the detector was found to be almost ideal. Moving away from the low-drive limit, we observed oscillations in the photocurrent, which we attribute to Rabi oscillations. We have also found an approximate and compelling analytical solution for the large-drive photocurrent by employing boundary-layer theory, capturing the qualitative behavior of the exact numerical solutions for the photocurrent. 

Our results pave the way for a better theoretical understanding of single microwave pulse photodetection inside a DQD-resonator system and could inspire future experiments in this avenue. Real-time measurements of electrical currents on the magnitudes and time-scales relevant in our system are very challenging. A detailed, quantitative investigation of the experimental approaches for such measurements is thus outside the scope of the present paper. Qualitatively, one could however imagine several ways in which this could be done. First, real-time detection of individual photoelectron tunnel events have been reported in Ref.~\cite{Haldar4}, albeit with a very low efficiency. Since the microwave pulses are identical and arrive at the resonator at well controlled times, an improved detector efficiency would enable a time-dependent photocurrent to be reconstructed directly from averaging over many time-resolved electron tunnel events. Second, the physical information in the time-resolved current is typically also reflected in the finite-frequency photocurrent noise,  which is experimentally accessible. A theoretical investigation of the finite-frequency photocurrent noise could be performed along the lines of our earlier paper, Ref.~\cite{Zenelaj}. Third, one could also imagine interferometric set-ups as in Ref.~\cite{Ferraro2013}, where the electronic current pulse is sent onto an electronic Mach-Zehnder interferometer with controllable arm-length difference. This would allow for a systematic investigation of the time-dependent photocurrent via the output currents from the interferometer. On the theoretical side, expressing the photodetection as a Volterra series may enable an analytic understanding of photodetection beyond weak drives. In addition, the development of a Wigner-function formalism for the outgoing photocurrent pulse may shed further insight into the photodetection process. Other intriguing avenues for research include examining different types of drives, such as non-classical states of light, as well as the potential to generate said states and investigate their statistical properties \cite{maisi,eichler,tineke,segal}.

\section*{Acknowledgments}
This work was supported by the Knut and Alice Wallenberg
Foundation through the Wallenberg Center for Quantum Technology (WACQT). P.P.P. acknowledges funding from the Swiss National Science Foundation (Eccellenza Professorial Fellowship PCEFP2\_194268).

\appendix
\section{Photocurrent in the low-drive limit}\label{AA}
To compute the photocurrent in Eq.~\eqref{pc}, we need to find the occupations of the excited state $p_e$ and the empty state $p_0$. These quantities are found using the master equation \eqref{LME} together with the equality $\expval{x}=\text{Tr}\{x\rho(t)\}$. This leads to a set of coupled equations of motion. Expanding this set of equations to lowest order in the drive gives
\begin{widetext}
\begin{align}
    \partial_t p_0(t)&\! \begin{aligned}[t]&=-\Gamma_{g0}p_0(t)+\Gamma_{0e}p_e(t), \end{aligned} \label{p0}\\[0.8em]
    \partial_t p_e(t)&\! \begin{aligned}[t]&=-(\Gamma_{0e}+\gamma_-)p_e(t)+2 g \text{Im}\expval{a\sigma_+}(t), \end{aligned} \label{pe}\\[0.8em]
    \partial_t \text{Im}\expval{a\sigma_+}(t)&\! \begin{aligned}[t]&=-\frac{1}{2}(\tilde{\Gamma}+\kappa)\text{Im}\expval{a\sigma_+}(t)-g p_e(t)-\frac{\sqrt{\kappa_{\text{in}}}}{2}[f(t) \expval{\sigma_+}(t)+f^*(t) \expval{\sigma_-}(t)] +g\expval{a^{\dagger}a}(t), \end{aligned} \label{ima}\\[0.8em]
    \partial_t \expval{\sigma_-}(t)&\! \begin{aligned}[t]&=-\frac{\tilde{\Gamma}}{2} \expval{\sigma_-}(t)-ig\expval{a}(t), \end{aligned} \label{sm}\\[0.8em]
    \partial_t\expval{a}(t)&\! \begin{aligned}[t]&=-\frac{\kappa}{2}\expval{a}(t)-i\sqrt{\kappa_{\text{in}}}f(t)-ig \expval{\sigma_-}(t), \end{aligned} \label{at}\\[0.8em]
    \partial_t\expval{a^{\dagger}a}(t)&\! \begin{aligned}[t]&=-\kappa\expval{a^{\dagger}a}(t)-2g \text{Im}\expval{a\sigma_+}(t)+i\sqrt{\kappa_{\text{in}}}\left[f^*(t)\expval{a}(t)-f(t)\expval{a^{\dagger}}(t)\right].\end{aligned} \label{adat}
\end{align}
The equations are most readily solved in Fourier space. In what follows, we use these definitions for the Fourier transform
\begin{equation}
\tilde{f}(\omega)=\frac{1}{\sqrt{2\pi}}\int dt \, f(t) e^{i\omega t}, \hspace{2cm}
\tilde{f}(\omega,\omega')=\frac{1}{2\pi}\int dtdt'\, f(t,t')e^{i(\omega t+\omega' t')}.
\end{equation}
We plug the solutions for $p_e$ and $p_0$ in Fourier space into the Fourier transform of the current in Eq.~\eqref{pc}, which is given by $\tilde{I}(\omega)/e=\Gamma_\text{R,out} p_e(\omega)-\Gamma_\text{R,in} p_0(\omega)$. The photocurrent is then found to be of the following form
\begin{equation}\label{curfs}
     \tilde{I}(\omega)/e=\frac{1}{\sqrt{2\pi}} \int  \frac{d\omega'}{2\pi} \tilde{h}(\omega-\omega',\omega')\tilde{f}(\omega-\omega')\tilde{f}^*(\omega'),
\end{equation}
where $\tilde{f}^*(\omega)$ is the Fourier transform of $f^*(t)$ and $\tilde{h}(\omega,\omega')=\tilde{g}(\omega,\omega')+\tilde{g}(\omega',\omega)$, with

\begin{equation}\label{gg}
    \tilde{g}(\omega,\omega')=\frac{16\pi g^2\kappa_\text{in}[-(\tilde{\Gamma}+\kappa)+i(\omega+3\omega')]\{\Gamma_{0e}\Gamma_\text{R,in}-\Gamma_\text{R,out}[\Gamma_{g0}-i(\omega+\omega')]\}}{[\Gamma_{g0}-i(\omega+\omega')][4g^2+(\tilde{\Gamma}-2i\omega')(\kappa-2i\omega')]\Lambda},
\end{equation}
where
\begin{equation}\label{Lambda}
    \Lambda= 4g^2[\Gamma_{0e}+\gamma_-+\kappa-2i(\omega+\omega')]+[\tilde{\Gamma}+\kappa-2i(\omega+\omega')][(\Gamma_{0e}+\gamma_--i(\omega+\omega')][\kappa-i(\omega+\omega')].
\end{equation}
By plugging in all the requirements for the ideal-detector limit except the unit cooperativity, we find
\begin{equation}\label{egg}
    \tilde{g}(\omega,\omega')=\frac{16\pi g^2\Gamma_\text{R}\kappa[\Gamma_\text{R}+\kappa-i(\omega+3\omega')]}{[\Gamma_\text{R}+\kappa-2 i(\omega+\omega')][4g^2+(\Gamma_{\rm R}-2i\omega')(\kappa-2i\omega')]\{4g^2+[\Gamma_{\rm R}-i(\omega+\omega')][\kappa-i(\omega+\omega')]\}}. 
\end{equation}

In the full ideal-detector limit, Eq.~\eqref{gg} becomes
\begin{equation}\label{filterfourier}
    \tilde{g}(\omega,\omega')=\frac{2\pi\Gamma_\text{R}^2\kappa^2[\Gamma_\text{R}+\kappa-i(\omega+3\omega')]}{[\Gamma_\text{R}+\kappa-2 i(\omega+\omega')][\Gamma_\text{R}\kappa-i\omega'(\Gamma_\text{R}+\kappa)-2\omega'^2]\{2\Gamma_\text{R}\kappa-i(\omega+\omega')[\Gamma_\text{R}+\kappa-i(\omega+\omega')]\}}.  
\end{equation}
Taking the Fourier transform of Eq.~\eqref{curfs} and replacing $\omega \to \omega+\omega'$, we get
\begin{equation}\label{ftiw}
    I(t)/e=\frac{1}{(2\pi)^2}\int d\omega d\omega' \, e^{-i(\omega+\omega')t} \tilde{h}(\omega,\omega')\tilde{f}(\omega)\tilde{f}^*(\omega')= \frac{1}{2\pi}\int_0^{\infty} d\tau d\tau'h(\tau,\tau')f(t-\tau)f^*(t-\tau'), 
\end{equation}
where $h(\tau,\tau')$ is the Fourier transform of $\tilde{h}(\omega,\omega').$ We note that $h(\tau,\tau') =h(\tau',\tau)$. The integral in Eq.~\eqref{ftiw} only runs over positive times. This is because the function $\tilde{g}(\omega,\omega')$ in Eq.~\eqref{gg} only has poles in the lower half plane, which results in $h(\tau,\tau')$ vanishing at negative times. We also note that $\tilde{h}(\omega,\omega')=\tilde{h}^*(-\omega,-\omega'),$ and thus $h(\tau,\tau')=h^*(\tau,\tau')$.
We now have all the tools to introduce the Wigner function for the detector
\begin{equation}\label{wigner}
    W_\text{d}(t,\omega)=\frac{1}{2\pi}\int d\xi e^{-i\xi\omega} h(t+\xi/2,t-\xi/2) = \frac{1}{2\pi}\int d\chi e^{i\chi t}\tilde{h}(\omega+\chi/2,\omega-\chi/2).
\end{equation}
The second equation in \eqref{wigner} is used to compute the detector Wigner function in Eq.~\eqref{idealwigner} in the main text by plugging Eq.~\eqref{egg} into the definition of $\tilde{h}(\omega
,\omega')$.
In addition, we define a Wigner function for the drive
\begin{equation}\label{wigdri}
    W_\text f (t,\omega) = \frac{1}{2\pi} \int d\xi e^{-i\xi\omega} f(t+\xi/2)f^*(t-\xi/2).
\end{equation}
This allows us to write the photocurrent as 
\begin{equation}\label{curwiggFour}
    I(t)/e = \int d\tau d\omega W_\text{d}(\tau,\omega) W_\text f (t-\tau,\omega). 
\end{equation}
\section{Photocurrent expressions in the ideal-detector limit}\label{AB}
In the frequency domain, we find an analytical expression for the photocurrent in the ideal-detector limit when the system is driven by a Gaussian pulse by plugging Eq.~\eqref{gaussf} into Eq.~\eqref{curfs}. We find
\begin{equation}
    \tilde{I}(\omega)/e=\frac{\sqrt{\pi}\bar{N}\Gamma_\text{R}^2\kappa^2\sigma e^{-\frac{\sigma^2[(2\omega-iy)^2+4\omega^2]}{16}}\{e^{\frac{i\alpha\sigma^2(\Gamma_\text{R}+\kappa-2i\omega)}{4}}(iy+2\omega)\text{erfc}[\frac{\sigma(x-2i\omega)}{4}]-(ix+2\omega)\text{erfc}[\frac{\sigma(y+2i\omega)}{4}]\}}{\alpha(\Gamma_\text{R}+\kappa-2i\omega)[2\Gamma_\text{R}\kappa-i\omega(\Gamma_\text{R}+\kappa)-\omega^2]},
\end{equation}
where $x=\Gamma_\text{R}+\kappa+i\alpha$, $y=\Gamma_\text{R}+\kappa-i\alpha$ and $\text{erfc}(z)$ gives the complementary error function. 
For the case of a Lorentzian pulse drive, we plug Eq.~\eqref{lorentzf} into Eq.~\eqref{curfs} and find

\begin{equation}
\begin{split}
    \tilde{I}(\omega)/e=& \frac{\bar N \Gamma_\text{R}^2 \kappa^2 \sigma}{\alpha(\Gamma_\text{R}+\kappa-2i\omega)[2\Gamma_\text{R}\kappa-i\omega(\Gamma_\text{R}+\kappa)-\omega^2]} \\
    &\times\Big\{ e^{\frac{\sigma}{2}(iy+2\omega)}(x-2i\omega)\{ \Gamma(0,\sigma(iy+2\omega)/2)-[\Gamma(0,\sigma(iy+2\omega)/2)-\Gamma(0,i\sigma y/2)]\Theta(-\omega)\} \\
    &- e^{-\frac{\sigma}{2}(iy+2\omega)}(x-2i\omega)\{ \Gamma(0,-\sigma(iy+2\omega)/2)-[\Gamma(0,-\sigma(iy+2\omega)/2)-\Gamma(0,-i\sigma y/2)]\Theta(\omega)\} \\
    &-e^{\frac{\sigma}{2}(ix+2\omega)}(y-2i\omega)\big\{\Gamma(0,\sigma(ix+2\omega)/2)+[\Gamma(0,i\sigma x/2)-\Gamma(0,\sigma(ix+2\omega)/2)]\Theta(-\omega) \\
    &- e^{-\sigma(ix+2\omega)}\{\Gamma(0,-\sigma(ix+2\omega)/2)-[\Gamma(0,-\sigma(ix+2\omega)/2)-\Gamma(0,-i\sigma x/2)]\Theta(\omega)\} \big\}\Big\},
\end{split}
\end{equation}
where $\Gamma(a,z)$ is the incomplete Gamma function.
The time-dependent photocurrent in Fig.~\ref{fig:c_g} is then found by inverse Fourier transformation. 
\end{widetext}

\section{Properties of the Wigner functions}\label{AD}
In this appendix, we derive properties of the Wigner function including the bound on its absolute value in Eq.~\eqref{eq:wignerbound} as well as the time-energy uncertainty relation in Eq.~\eqref{uncertain}. To this end, we are going to use analogies to the phase space Wigner function used in quantum mechanics making use of the following properties of the function $h(\tau,\tau')$ that are proven in App.~\ref{AA}
\begin{equation}
\label{eq:hprop}
    h(\tau,\tau') = h(\tau',\tau),\hspace{.25cm}h^*(\tau,\tau') =h(\tau,\tau').
\end{equation}

\subsection{Analogy to phase space Wigner function}
We may interpret $h$ as an operator on the vector space of square-integrable functions acting on a function $u(\tau)$ as 
\begin{equation}
   (hu)(\tau) \equiv \int d\tau' \, h(\tau,\tau')u(\tau').
\end{equation}
The inner product on this space is defined in the usual manner
\begin{equation}
    \label{eq:inner}
    \langle v,u\rangle =  \int d\tau \, v^*(\tau)u(\tau).
\end{equation}
With Eqs.~\eqref{eq:hprop}-\eqref{eq:inner}, one may show that
\begin{equation}
    \langle v, hu\rangle = \langle hv, u\rangle,
\end{equation}
implying that $h$ is a Hermitian operator. We now introduce the normalized function
\begin{eqnarray}
    \label{eq:hbar}
    \bar{h}(\tau,\tau') = \frac{h(\tau,\tau')}{\int dx \, h(x,x)}.
\end{eqnarray}
We may use the spectral theorem to write 
\begin{equation}
    \label{eq:hspec}
    \bar{h}(\tau,\tau') = \sum_j p_j u_j(\tau) u_j^*(\tau'),
\end{equation}
where $u_j$ denote the orthonormal eigenvectors of $h$, i.e.,
\begin{equation}
    \int d\tau' \, \bar{h}(\tau,\tau')u_j(\tau') = p_ju_j(\tau),
\end{equation}
with
\begin{equation}
   \int d\tau \, u_j^*(\tau)u_k(\tau) = \delta_{j,k}. 
\end{equation}
The normalization in Eq.~\eqref{eq:hbar} ensures that $\bar{h}$ has unit trace, i.e.,
\begin{equation}
    \sum_j p_j = 1.
\end{equation}

We note that there is an analogy between $\bar{h}(\tau,\tau')$ and the elements of a density matrix in quantum mechanics, which can be written as
\begin{equation}
    \label{eq:analrho}
    \langle x |\hat{\rho}|x'\rangle = \rho(x,x') = \sum_j p_j\psi_j(x)\psi_j^*(x').
\end{equation}
To make the analogy complete, we need to require that $p_j\geq 0$. This is ensured by assuming that $h(\tau,\tau')$ is a positive semi-definite function. From Eq.~\eqref{pcfilter}, we may infer that this is equivalent to assuming that the photocurrent $I(t)/e$ is non-negative at all times and for all drive amplitudes. To see this, we note that $h(\tau,\tau') = 0$ if either $\tau<0$ or $\tau'<0$. Equation \eqref{pcfilter} may thus be written as
\begin{equation}
  I(t)/e = \int d\tau d\tau' \, s^*(\tau)\bar{h}(\tau,\tau') s(\tau') \geq 0,
\end{equation}
with $s(\tau) = \sqrt{\int dx \, h(x,x)}f^*(t-\tau)/\sqrt{2\pi}$, which is the defining property of a positive semi-definite function. We note that for our detector, the current may become negative due to back-tunneling from the drain lead. The properties discussed in this appendix cannot be applied to this scenario.

With the analogy between $\bar{h}$ and a quantum-mechanical density matrix at hand, we may use standard properties of quantum-mechanical Wigner functions. We focus on two properties. First, we provide a bound on maximal and minimal values for $h$, and second, we prove the Heisenberg-Gabor limit in analogy to the Heisenberg uncertainty relation.

\subsection{Bounding the Wigner function}
With the results of the last section, we may prove Eq.~\eqref{eq:wignerbound} in complete analogy to the bound on the phase-space Wigner function, see for instance Ref.~\cite{curtright:2012}
\begin{equation}
\begin{aligned}
    &\frac{|W(t,\omega)|}{\bar{W}_{\rm d}} = \frac{1}{2\pi}\left|\int_{-\infty}^\infty d\xi e^{-i\xi\omega}\bar{h}(t+\xi/2,t-\xi/2) \right|\\&=\frac{1}{2\pi}\left|\sum_j p_j\int_{-\infty}^\infty d\xi e^{-i\xi\omega}u_j(t+\xi/2)u_j^*(t-\xi/2) \right|\\&
    =\frac{1}{2\pi}\left|\sum_j p_j\left\langle  u_j^*(t-\xi/2),e^{-i\xi\omega}u_j(t+\xi/2)\right\rangle \right|\\&
    \leq \frac{1}{2\pi}\sum_j p_j\left|\left\langle  u_j^*(t-\xi/2),e^{-i\xi\omega}u_j(t+\xi/2)\right\rangle \right|\\&
    \leq \frac{1}{2\pi}\sum_j p_j\langle u_j(\xi/2),u_j(\xi/2)\rangle\\&
   =\frac{1}{\pi}.
    \end{aligned}
\end{equation}
Here the first inequality is simply the triangle inequality, while the second inequality is the Cauchy-Schwarz inequality. In addition, we used the identities
\begin{equation}
    \langle e^{i\varphi} f,e^{i\varphi}g\rangle = \langle f,g\rangle,
\end{equation}
for any $\varphi\in \mathbb{R}$, and
\begin{equation}
    \langle u_j(t\pm\xi/2),u_j(t\pm\xi/2)\rangle = \langle u_j(\xi/2),u_j(\xi/2)\rangle,
\end{equation}
where the inner product is understood to be over the variable $\xi$.

\subsection{The Heisenberg-Gabor limit}
To prove the Heisenberg-Gabor limit \cite{Gabor1946} in Eq.~\eqref{uncertain}, we introduce the moments of the normalized Wigner function as
\begin{equation}
    \label{eq:momwig}
    \langle x \rangle=\int dt  d\omega \, x W_\text d(t,\omega)/ \bar W_\text d.
\end{equation}
From Eq.~\eqref{eq:hspec} we may infer that $\langle x\rangle = \sum_jp_j\langle x\rangle_j$. We thus find
\begin{equation}
\label{eq:deltafirst}
\begin{aligned}
    \Delta t\Delta\omega &= \sqrt{\sum_{j}p_j\langle (t-\langle t\rangle)^2\rangle_j\sum_kp_k\langle (\omega-\langle \omega\rangle)^2\rangle_k}\\&\geq \sum_j p_j\sqrt{\langle (t-\langle t\rangle)^2\rangle_j\langle (\omega-\langle \omega\rangle)^2\rangle_j},
    \end{aligned}
\end{equation}
where we employed the Cauchy-Schwarz inequality. Next we write
\begin{equation}
    \label{eq:innerts}
    \begin{aligned}
    &\langle (t-\langle t\rangle)^2\rangle_j = \int dt u_j^*(t)(t-\langle t\rangle)^2u_j(t)\\&= \langle (t-\langle t\rangle)u_j(t), (t-\langle t\rangle)u_j(t)\rangle,
    \end{aligned}
\end{equation}
where we note the difference between the averages $\langle x\rangle$ and $\langle f,g\rangle$, which are defined in Eqs.~\eqref{eq:momwig} and \eqref{eq:inner} respectively.

Similarly to how the momentum operator can be written as a derivative of position, we may obtain the frequency moments using time derivatives as
\begin{equation}
    \label{eq:innerws}
     \begin{aligned}
    &\langle (\omega-\langle \omega\rangle)^2\rangle_j = \int dt u_j^*(t)(-i\partial_t-\langle \omega\rangle)^2u_j(t)\\&= \langle (-i\partial_t-\langle \omega\rangle)u_j(t), (-i\partial_t-\langle \omega\rangle)u_j(t)\rangle,
    \end{aligned}
\end{equation}
where the first equality can be derived from Eq.~\eqref{wigner} and the last equality can be proven using integration by parts. We may now use the Cauchy-Schwarz inequality to write
\begin{equation}
\begin{aligned}
   &\sqrt{\langle r(t), r(t)\rangle \langle s(t),s(t)\rangle}\geq |\langle r(t),  s(t)\rangle|\geq \left|{\rm Im}\{\langle r(t),  s(t)\rangle\}\right|.
\end{aligned}
\end{equation}
With $r(t) = (t-\langle t\rangle)u_j(t)$ and $s(t) = (-i\partial_t-\langle \omega\rangle)u_j(t)$, we find
\begin{equation}
    \sqrt{\langle (t-\langle t\rangle)^2\rangle_j\langle (\omega-\langle \omega\rangle)^2\rangle_j}\geq \frac{1}{2},
\end{equation}
where we again used integration by parts. Inserting the last inequality into Eq.~\eqref{eq:deltafirst}, we recover the Heisenberg-Gabor limit in Eq.~\eqref{uncertain}.
\section{Wigner function symmetry in $\Gamma_{\rm R}$ and $\kappa$}\label{AD2}
Here we consider the equations of motion for weak drives given in Eqs.~\eqref{p0}-\eqref{adat}, with the additional simplification $\gamma_\phi=\gamma_-=0$ as well as $\Omega\gg t_c$. In this case, we find
\begin{equation}
\begin{aligned}
    \partial_t \expval{\sigma_-}(t)&=-\frac{\Gamma_\text R}{2} \expval{\sigma_-}(t)-ig\expval{a}(t), \\
    \partial_t\expval{a}(t)&=-\frac{\kappa}{2}\expval{a}(t)-i\sqrt{\kappa_{\text{in}}}f(t)-ig \expval{\sigma_-}(t).
    \end{aligned}
\end{equation}
The other relevant equations factorize, i.e., $\langle a^\dagger a\rangle = \langle a^\dagger\rangle\langle a\rangle$, $\langle a^\dagger \sigma_-\rangle = \langle a^\dagger \rangle\langle\sigma_-\rangle$, and $p_e = \langle \sigma_+\rangle\langle \sigma_-\rangle$. In this case, the photodetector behaves just like two coupled cavities, one of which is weakly driven. This can be understood by appreciating that the empty state of the DQD becomes irrelevant, as the state always returns to the ground state before the next photon arrives due to the weak drive. The ground and excited states then map onto the lowest energy states of a harmonic oscillator, with the higher states being irrelevant due to the weak drive.

The photocurrent reduces to [c.f.~Eq.~\eqref{pc}] $I(t) = \Gamma_\text R \langle \sigma_+\sigma_-\rangle$, which is nothing else than the fictitional photons leaking out of the second cavity with linewidth $\Gamma_\text R$. The photon-to-electron conversion is thus mathematically equivalent to the transmission of photons through two coupled cavities. Exchanging $\kappa \leftrightarrow \Gamma_\text R$ amounts to applying the drive to the cavity with linewidth $\Gamma_\text R$ and considering the photon flux out of the cavity with linewidth $\kappa$ as the current $I(t)$.

Further insight may be obtained by defining two new operators $q_{\pm}$ and their Hermitian conjugates $q_{\pm}^\dagger$, as
\begin{equation} 
q_{\pm}=\frac{1}{\sqrt{2}}\left(a\pm \sigma_-\right), \hspace{0.5 cm} q_{\pm}^{\dagger}=\frac{1}{\sqrt{2}}\left(a^{\dagger} \pm \sigma_+\right).
\end{equation}
For the resonant DQD-resonator coupling case, these operators create the two first excited states of the Jaynes-Cummings Hamiltionian, when acting on the DQD-resonator ground state $|g0\rangle$. We can then rewrite the equations of motion for $\langle a \rangle$ and $\langle \sigma_- \rangle$, considering here only the ideal limit, as  
\begin{eqnarray} 
\partial_t \langle q_+ \rangle&=&-\left(\frac{\Gamma_{\rm R}+\kappa}{4}+ig\right) \langle q_+ \rangle-(\Gamma_{\rm R}-\kappa) \langle q_- \rangle-\frac{if}{\sqrt{2}}, \nonumber \\
\partial_t \langle q_- \rangle&=&-\left(\frac{\Gamma_{\rm R}+\kappa}{4}-ig\right) \langle q_- \rangle-(\Gamma_{\rm R}-\kappa) \langle q_+ \rangle+\frac{if}{\sqrt{2}}. \nonumber \\
\end{eqnarray}
From these equations we can draw the following conclusions: 
\begin{itemize}
\item The equations describe two coherently coupled and driven electron-photon states. The states are the first excited Jaynes-Cummings ladder states, higher excited states do not contribute in the weak drive limit. 
\item The states have a decay rate $\Gamma_{\rm R}+\kappa$, due to the coupling of the electron-photon system to the electronic and photonic contacts/reservoirs. The energy difference between the two states is given by $2g$. 
\item The coupling between the probabilities and the superpositions of the states is proportional to $\Gamma_{\rm R}-\kappa$. Thus, for the special case of $\Gamma_{\rm R}=\kappa=\Gamma$ the two states decouple.
\item The probability $p_e$ determining the electrical current is found to be proportional to the drive amplitude squared, i.e. $f^2 \kappa$. However, the electrical current, see Eq. (\ref{pc}), is in the considered limit proportional to $\Gamma_{\rm R}p_e$, that is $\propto\Gamma_{\rm R}\kappa$. 
\end{itemize}

Taken together, this discussion illustrates that the electron and the photon aspects of the system enter symmetrically in the final expressions for the electrical current. Or, in other words, the Wigner function is invariant under $\kappa \leftrightarrow \Gamma_{\rm R}$.

\section{Approximate solution, boundary-layer theory}\label{AE}
Here we discuss the derivation of the boundary-layer theory (BLT) approximation for the photocurrent, c.f. Eq.~\eqref{BLT}. The starting point is the Rabi equations in Eq.~\eqref{Rabi} for the DQD density matrix components, keeping in mind that the photocurrent is given by $I(t)=e\Gamma_{\rm R}p_e(t)$. Note that the difference from the photocurrent introduced in Eq.~\eqref{pc} is due to the small interdot tunneling requirement in Tab.~\ref{tab:ideal}, which leads to $\theta \to 0$. As pointed out in the main text, at times much shorter than the inverse decay rate, $t\ll 1/\Gamma_{\rm R}$, the DQD evolution is fully coherent and the photocurrent is given by the area law expression in Eq.~\eqref{arealaw}, 

\begin{equation}
I_\text{in}(t)= e\Gamma_{\rm R}\sin^2\left[\mathcal{A}(t)\right] , \hspace{0.3cm} \mathcal{A}(t) =\int_{-\infty}^{t} dt'\Omega(t').
\end{equation}
where we here introduced the subscript ``in'', denoting the \textit{inner} solution according to the nomenclature of BLT. 

Focusing on the regime of weak decay, $\Gamma_{\rm R} \ll \kappa$, in the limit of times much longer than the resonator photon decay time, $t\gg 1/\kappa$,  the resonator is empty. As pointed out in the main text, the photocurrent is then given directly via the equation for $p_e(t)$ in Eq.~\eqref{Rabi}, as

\begin{equation}
I_\text{out}(t)=ce\Gamma_{\rm R}e^{-t\Gamma_{\rm R}},
\end{equation}
where $c$ is a constant and we added the subscript ``out'', denoting the \textit{outer} solution. In order to obtain the BLT expression for the photocurrent, describing also times $1/\kappa < t < 1/\Gamma_{\rm R}$, the first step is to determine the constant $c$ by requiring that the asymptotic photocurrent expressions are equal, as (assuming a pulse that starts around $t=0$)

\begin{equation}
I_\text{in}(\infty)=I_\text{out}(0) \hspace{0.3 cm} \Rightarrow \hspace{0.3 cm} c=\sin^2\left[\mathcal{A}(\infty)\right].
\end{equation}%
As a second step, an approximate, BLT photocurrent expression is obtained by adding the inner and outer solutions and then subtracting their common asymptotic value, as

\begin{eqnarray}
I_\text{BLT}(t)&=&I_\text{in}(t)+I_\text{out}(t)- e\Gamma_{\rm R}\sin^2\left[\mathcal{A}(\infty)\right] \\ \nonumber
&=&  e\Gamma_{\rm R}\left\{\sin^2\left[\mathcal{A}(t)\right]+\sin^2\left[\mathcal{A}(\infty)\right]\left(e^{-t\Gamma_{\rm R}}-1\right)\right\},
\label{BLTa}
\end{eqnarray}
Finally, by writing the expression in the curly brackets in Eq.~\eqref{BLTa} as $\sin^2\left[\mathcal{A}(t)\right]e^{-t\Gamma_{\rm R}}+\left\{\sin^2\left[\mathcal{A}(\infty)\right]-\sin^2\left[\mathcal{A}(t)\right]\right\}\left(e^{-t\Gamma_{\rm R}}-1\right)$, we see that the second term in this expression is much smaller than the first one for all relevant $t$ and can thus simply be dropped. This then gives the photocurrent
\begin{equation}
I_\text{BLT}(t)=e\Gamma_{\rm R}\sin^2\left[\mathcal{A}(t)\right]e^{-t\Gamma_{\rm R}},
\end{equation}
which is the expression presented in Eq.~\eqref{BLT} in the main text.

\section{Exact result, impulse response}\label{AF}
Here we present the derivation of the formal solution for the Rabi equations in Eq.~\eqref{Rabi} for a $\delta$-function pulse, i.e., the impulse response $f(t) = A\delta(t)$, with $A$ being a real constant. As a starting point, using Eq.~\eqref{amp} for the resonator amplitude $\langle a \rangle(t)$ with the initial condition $\langle a \rangle(-\infty)=0$, a $\delta$-function input gives 
\begin{equation}
\langle a \rangle(t) = -i\sqrt{\kappa}Ae^{-\kappa t/2}, \hspace{0.3 cm} t\geq 0.
\label{aeq}
\end{equation}
The amplitude is purely imaginary, $\langle a^{\dagger} \rangle(t)=-\langle a \rangle(t)$, and exponentially decaying with time after a step onset at $t=0$. Accounting for the imaginary $\langle a \rangle(t)$ and making use of the normalization $p_e(t)+p_g(t)=1$, we can reduce the Rabi equations to two coupled linear differential equations for $p_e(t)$ and $\expval{\sigma_x}(t)=\expval{\sigma_-+\sigma_+}(t)$, as
\begin{equation}\label{effrabi} 
\begin{split}
\partial_t p_e(t) &= -\Gamma_{\rm R} p_e(t) -ig\langle a \rangle(t)\sigma_x(t) , \\
\partial_t\expval{\sigma_x}(t) &=-\frac{\Gamma_{\rm R}}{2}\sigma_x(t)+ig\langle a \rangle(t)\left[2p_e(t)-1\right], 
\end{split}
\end{equation}
here writing out $\Omega(t)=-\Omega^*(t)=-g\langle a \rangle(t)$ for clarity. To obtain an equation for $p_e(t)$ only, we take the derivative of the first equation in Eq.~\eqref{effrabi} and then insert the expression for $\partial_t \expval{\sigma_x}(t)$ from the second equation and $\expval{\sigma_x}(t)=i[\Gamma p_e(t)+\dot p_e(t)]/g$ from the first, giving 

\begin{widetext}
\begin{equation}
\partial_t^2 p_e(t) = -\left[\frac{3\Gamma_{\rm R}}{2}-\frac{\langle \dot{a}\rangle (t)}{\langle a \rangle (t)}\right]\dot p_e(t)-\Gamma_{\rm R} \left[\frac{\Gamma_{\rm R}}{2}-\frac{\langle \dot{a}\rangle (t)}{\langle a\rangle (t)}\right]p_e(t) + g^2\langle a\rangle^2(t)[2p_e(t)-1].
\end{equation}
Inserting the expression for $\langle a \rangle(t)$ in Eq.~\eqref{aeq}, making use of that $\langle \dot{a}\rangle (t)/\langle a \rangle= -\kappa/2$ and rearranging, we can write the equation for $p_e(t)$ in the form 
\begin{equation}
\partial_t^2 p_e(t)+\frac{3\Gamma_{\rm R}+\kappa}{2}\dot p_e(t)+\left[\frac{\Gamma_{\rm R}(\Gamma_{\rm R}+\kappa)}{2}+4g^2\kappa A^2e^{-t\kappa} \right]p_e(t) =2g^2\kappa A^2 e^{-t\kappa}.
\end{equation}
To proceed, we write $p_e(t)=y(t)e^{-(3\Gamma_{\rm R}+\kappa)t/4}$ and change variables as $x=4gAe^{-\kappa t/2}/\sqrt{\kappa}$, giving the equation for $y(x)$ as

\begin{equation}
x^2\frac{d^2y(x)}{dx^2}+x\frac{dy(x)}{dx}+(x^2-\nu^2)y(x)=\mu^{2+3\nu} x^{-3\nu},
\end{equation}
where we introduced $\nu=(\Gamma_{\rm R}/\kappa-1)/2$ and $\mu=4gA/\sqrt{\kappa}$. We note that putting the right hand side equal to zero, the resulting homogenous equation is Bessel's equation, with solutions $J_\nu(x)$ and $Y_\nu(x)$. The general solution is given by the sum of the homogenous solution, $y_h(x)=C_1J_\nu(x)+C_2Y_\nu(x)$ with $C_1,C_2$ constants, and a particular solution
\begin{equation}
\begin{split}
y_p(x)&=\frac{\pi}{2}\mu^{2+3\nu} J_\nu(x)\int dx Y_\nu(x)x^{1-3\nu}  \\
&-\frac{\pi}{2}\mu^{2+3\nu}Y_\nu(x)\int dx J_\nu(x) x^{1-3\nu},
\end{split}
\end{equation}
where we have used that the Wronskian $(dY_\nu(x)/dx)J_\nu(x)-(dJ_\nu(x)/dx)Y_\nu(x)=2/(\pi x)$. Performing the integrals and implementing the initial conditions $y(\mu)=dy(x)/dx|_{x=\mu}=0$, the full solution for $y(x)$ can be written in the arguably not very transparent way, as 

\begin{eqnarray}
y(x)&=&\frac{\pi 2^{-3+\nu}\mu^{2-\nu}x^{-4\nu}J_\nu(x)\csc(\pi\nu)}{\nu^2\Gamma(-\nu)} \left[\mu^{4\nu}{}_1F_2\left(-2\nu;1-2\nu,1-\nu;-\frac{x^2}{4}\right) \right. - \left. x^{4\nu}{}_1F_2\left(-2\nu;1-2\nu,1-\nu;-\frac{\mu^2}{4}\right)\right] \nonumber \\
&+& \frac{2^{-2-\nu}\mu^{2+\nu}x^{-2\nu}J_{-\nu}(x)\Gamma(-\nu)}{\nu} \left[x^{2\nu}{}_1F_2\left(-\nu;1-\nu,1+\nu;-\frac{\mu^2}{4}\right) \right.
 - \left. \mu^{2\nu}{}_1F_2\left(-\nu;1-\nu,1+\nu;-\frac{x^2}{4}\right)\right],
\end{eqnarray}
where $_pF_q$ is the hypergeometric function and $\Gamma$ the Gamma function. From above we then have 
\begin{equation}
p_e(t)=e^{-(3\Gamma_{\rm R}+\kappa)t/4}y(4gAe^{-\kappa t/2}/\sqrt{\kappa}).
\end{equation}
\end{widetext}

\bibliography{main.bib}

\end{document}